\documentclass[preprint2]{aastex}
\usepackage{color}

\shorttitle{Classification of Microwave Zebra Patterns}
\shortauthors{B.L. Tan}

\begin{document}

\title{Statistics and Classification of the Microwave Zebra Patterns Associated with Solar Flares}

\author{Baolin Tan$^1$, Chengming Tan$^1$, Yin Zhang$^1$, H. M\'esz\'arosov\'a$^2$, M. Karlick\'y$^2$}
\affil{$^1$ Key Laboratory of Solar Activity, National
Astronomical Observatories of Chinese Academy of Sciences, Beijing
100012, Email: bltan@nao.cas.cn} \affil{$^2$ Astronomical
Institute of the Academy of Sciences of the Czech Republic,
Ondrejov 15165}

\begin{abstract}

The microwave zebra pattern (ZP) is the most interesting,
intriguing, and complex spectral structure frequently observed in
solar flares. A comprehensive statistical study will certainly
help us to understand the formation mechanism, which is not
exactly clear now. This work presents a comprehensive statistical
analysis on a big sample with 202 ZP events collected from
observations at the Chinese Solar Broadband Radio Spectrometer at
Huairou and the Ond\'rejov Radiospectrograph in Czech Republic at
frequencies of 1.00 - 7.60 GHz during 2000 - 2013. After
investigating the parameter properties of ZPs, such as the
occurrence in flare phase, frequency range, polarization degree,
duration, etc., we find that the variation of zebra stripe
frequency separation with respect to frequency is the best
indicator for a physical classification of ZPs. Microwave ZPs can
be classified into 3 types: equidistant ZP, variable-distant ZP,
and growing-distant ZP, possibly corresponding to mechanisms of
Bernstein wave model, whistler wave model, and double plasma
resonance model, respectively. This statistical classification may
help us to clarify the controversies between the existing various
theoretical models, and understand the physical processes in the
source regions.

\end{abstract}


\section{Introduction}

As we have known, zebra pattern (ZP) is a kind of spectral fine
structure superposed on the solar radio broadband type IV
continuum spectrogram, which consists of several almost parallel
and equidistant stripes. It is the most intriguing and interesting
fine structure on the dynamic spectra of solar radio observations,
especially at microwave frequency range, which may reveal the
original information of the solar flaring processes, such as the
magnetic fields and its configurations, particle acceleration, and
the plasma features in the source regions where the primary energy
releases. The nature of ZP structures has been a widely discussed
subject for more than 40 years. The historical development of
observations and various theoretical models is assembled in the
review of Chernov (2006), Zlotnik (2009), and so on (Rosenberg
1972, Kuijpers 1975, Zheleznyakov \& Zlotnik 1975, Chernov 1976,
1990, LaBelle et al. 2003, Kuznetsev 2005, Ledenev et al. 2006,
Tan 2010, Karlicky 2013). These models include:

(1) Bernstein wave (BW) model

It is the first model interpreting the formation of ZP structure
which proposed that all the stripes in a ZP structure are
generated from a small compact source, and the emission produces
from some nonlinear coupling processes between two Bernstein
waves, or Bernstein wave and other electrostatic upper hybrid
waves. The electrons with non-equilibrium distribution over
velocities perpendicular to the magnetic field are located in a
small source, where the plasma is weakly and uniformly magnetized
($f_{pe}\gg f_{ce}$). These electrons excite longitudinal
electrostatic waves at frequency of the sum of Bernstein modes
frequency $sf_{ce}$ and the upper hybrid frequency $f_{uh}$:
$f=f_{uh}+sf_{ce}\approx f_{pe}+sf_{ce}$. Here, $f_{pe}$ is the
electron plasma frequency, $f_{ce}$ the electron gyro-frequency,
$s$ is harmonics number. The BW excitation occurs in relatively
narrow frequency band. This model predicts the frequency
separation between the adjacent zebra stripes just as the electron
gyro-frequency: $\Delta f=f_{ce}$ (Rosenberg 1972, Chiuderi et al.
1973, Zheleznyakov \& Zlotnik 1975, Zaitsev \& Stepanov 1983), it
approximates a constant.

(2) Whistler wave (WW) model

This is an important model based on propagation of whistler wave
packets across or along the magnetic loop where the energetic
electrons generate Langmuir waves (Kuijpers 1975, Chernov 1976,
Maltseva \& Chernov 1989). The quasi-standing whistler packets can
be driven by loss-cone distribution of fast electrons in the
entire magnetic trap at some cyclotron resonance conditions. The
coupling of plasma Langmuir waves and whistler wave packets can
operate in different resonance conditions: when whistlers generate
at the normal Doppler cyclotron resonance they can escape along
the magnetic loop and yield fiber bursts
($f_{w}-\frac{k_{\parallel}}{2\pi}v_{\parallel}-f_{ce}=0$, $f_{w}$
is the whistler wave frequency, $k_{\parallel}$ is the whistler
wave number paralleled to magnetic field, $v_{\parallel}$ is the
fast electron velocity paralleled to magnetic field); when
whistlers generate at the anomalous Doppler cyclotron resonance
($f_{w}-\frac{k_{\parallel}}{2\pi}v_{\parallel}+f_{ce}=0$) under
large angles to the magnetic field they may form standing wave
packets in front of the shock wave, and when the group velocity of
whistlers is approximated to the shock velocity, a ZP structure
with slow oscillating frequency drift will appear. Each zebra
stripe corresponds to one propagating whistler wave packet. The
emission frequency at $f=f_{pe}+sf_{w}$, $f_{w}\approx 0.1 - 0.5
f_{ce}$ is the whistler wave frequency. Here, $f_{pe}\gg f_{ce}$.
The frequency separation $\Delta f$ of adjacent zebra stripes is
about 2 times of whistler wave frequency: $\Delta f\sim 2f_{w}$.
As $f_{w}$ varies in a small range, the whistler wave group
velocity peaks at frequency of $\frac{1}{4}f_{ce}$, therefore,
$\Delta f$ will vary around $\frac{1}{2}f_{ce}$.

(3) Double plasma resonance (DPR) model

The most developed heterogenous ZP model is called double plasma
resonance model (DPR model), which proposed that enhanced
excitation of plasma waves occurs at some resonance levels where
the upper hybrid frequency coincides with the harmonics of
electron gyro-frequency in the inhomogeneous flux tube (Pearlstein
et al. 1966, Zheleznyakov \& Zlotnik 1975, Berney \& Benz 1978,
Winglee \& Dulk 1986, Zlotnik et al. 2003, Yasnov \& Karlicky
2004, Kuznetsov \& Tsap 2007):
$f_{uh}=(f_{pe}^{2}+f_{ce}^{2})^{1/2}=sf_{ce}$. The emission
frequency is dominated not only by the electron gyro-frequency,
but also by plasma frequency. When the emission generates from the
coalescence of two excited plasma waves, the polarization may be
very weak, the emission frequency is $f\approx 2f_{pe}\approx
2sf_{ce}$, and the stripe frequency separation is $\Delta
f=\frac{2sf_{ce}H_{b}}{|sH_{b}-(s+1)H_{p}|}$. Here, $H_{b}$ and
$H_{p}$ are the scale heights of magnetic field and the plasma
density in the source regions, respectively. When the emission
generates from the coalescence of an excited plasma wave and a low
frequency electrostatic wave, the polarization will be strong, the
emission frequency is $f\approx f_{pe}\approx sf_{ce}$, and the
stripe frequency separation is $\Delta
f=\frac{sf_{ce}H_{b}}{|sH_{b}-(s+1)H_{p}|}$. In DPR model, $\Delta
f$ has a regular changing trend with a fairly large number of
stripes. Here, $\Delta f$ depends on $H_{b}$ and $H_{p}$, which
therein depends on the models of coronal magnetic field and plasma
density. For most models, we can deduce that $\Delta f$ will
increase with respect to the frequency.

(4) Propagating model

There are also some models proposed that ZP stripes could be
formed in the propagating processes after emitted from source
regions. The interference model suggests that ZP is formed from
some interference mechanisms in the propagating processes (B\'arta
\& Karlick\'y 2006, Ledenev et al. 2006). Some inhomogeneous
layers with small size may appear in source region, they can
change the emission into direct and reflected rays. When the
direct and reflected rays meet at some places, interference will
take place and produce ZP. This model needs a structure with great
number of discrete sources in small size, such structure may exist
in the current-carrying flaring plasma loop (Tan 2010) where the
tearing-mode instability forms a great number of magnetic islands
which may provide the main conditions for the interference
mechanism, similar to the crystal lattice. Very recently, Karlicky
(2013) proposed a new model that links ZP with propagating
compressive MHD waves. However, so far, the propagating model is
hard to predict the zebra stripe frequency separation, and the
relationship between the zebra parameters and the magnetic field
in the source region is also unknown.

Until now, the real formation of microwave ZPs is still
controversial. It is very difficult to interpret all observing
properties by using a unique existing model. It is meaningful to
make a classification of microwave ZPs. Possibly, the different
microwave ZPs may have different formation mechanism. Therefore, a
comprehensive statistical analysis of microwave ZPs is most
necessary.

In previous literatures, there are some statistical works on ZP
structures (Huang et al. 2008, 2010, Huang \& Tan 2012, Yu et al.
2012), but a physical classification is still missing. This work
will present a comprehensive statistical investigation on the
microwave ZPs at frequency of above 1000 MHz. Section 2 introduces
the observations and the composition of the statistical sample,
section 3 presents the statistical properties of ZP parameters. A
physical classification of ZPs is presented in section 4. Finally,
some conclusions discussions are summarized in section 5.

\section{Statistical Sample}

\subsection{Observation data}

In this work, the statistical sample in obtained from the
following two broadband solar radio spectrometers:

(1) The Chinese Solar Broadband Radio Spectrometers at Huairou
(SBRS/Huairou)

SBRS is an advanced solar radio telescope with super high cadence,
broad frequency bandwidth, and high frequency resolution, which
can distinguish the super fine structures from the spectrogram (Fu
et al. 1995, 2004, Yan et al. 2002). Its daily observational
window is 0:00-8:00 UT in winter seasons and 23:00-9:00 UT in
summer seasons. It includes 3 parts: 1.10 - 2.06 GHz (with the
antenna diameter of 7.0 m), 2.60 - 3.80 GHz (with the antenna
diameter of 3.2 m), and 5.20 - 7.60 GHz (share the same antenna of
the second part). The antenna points to the center of solar disk
automatically controlled by a computer. The spectrometer receives
the total flux of solar radio emission with dual circular
polarization (left- and right handed circular polarization, LCP
and RCP), and the dynamic range is 10 dB above quiet solar
background emission. The observation sensitivity is:
$S/S_{\bigodot}\leq 2\%$, here $S_{\bigodot}$ is the quiet solar
background emission. Our observation data includes:

During 2000-2003 and 2006-2008, 1.10-2.06 GHz with cadence of 5 ms
and frequency resolution of 4 MHz;

During 2004 - 2005, 1.10 - 1.34 GHz with cadence of 1.25 ms and
frequency resolution of 4 MHz;

During 2000 - 2013, 2.60 - 3.80 GHz with cadence of 8 ms and
frequency resolution of 10 MHz;

During 2000 - 2008, 5.20 - 7.60 GHz with cadence of 5 ms and
frequency resolution of 20 MHz.

(2) Ond\'rejov radiospectrograph in Czech Republic
(ORSC/Ond\'rejov).

ORSC is another broadband spectrometer located at Ond\'rejov,
Czech republic. It can receive solar radio total flux at
frequencies of 0.80 - 5.00 GHz during 2000 - 2013 (Jiricka et al.
1993). Its daily observational window is 7:00-16:00 UT in winter
seasons and 6:00-17:00 UT in summer seasons. Our observation data
includes:

During 2000 - 2005, 0.80 - 2.00 GHz with cadence of 100 ms and
frequency resolution of 5 MHz; 2.00 - 4.50 GHz with cadence of 100
ms and frequency resolution of 10 MHz;

During 2006 - 2013, 0.80 - 2.00 GHz with cadence of 10 ms and
frequency resolution of 5 MHz; 2.00 - 5.00 MHz with cadence of 100
ms and frequency resolution of 12 MHz.

SBRS/Huairou and ORSC/Ond\'rejov have a overlapping observational
window 7:00 - 8:00 UT in winter seasons and 6:00-9:00 UT in summer
seasons.

\subsection{Statistical Parameters}

It is necessary to make a clear definition of a ZP event. Here we
define a ZP event as an isolated spectral structure, which
consists of at least 2 almost-parallel stripes with approximately
equidistant separation and slowly frequency drifting rate, the
time gap between two such adjacent similar ZP events is at least
longer than the duration of each ZP event, and the frequency gap
between two such adjacent similar ZP events is at least wider than
the frequency range of each ZP event. Based on such definition, we
find that some flares have only one ZP event, while some flares
may accompany with several ZP events. By scrutinizing the
broadband spectrograms, we find that there are 154 ZP events in 27
flares observed by SBRS, and 49 ZP events in 13 flares observed by
ORSC during 2000 - 2013. There is only one ZP event observed
simultaneously by SBRS and ORSC in an M6.5 flare on 2013 April 11.
There are totally 202 ZP events in 40 flares, which are listed in
Table 1.

\begin{deluxetable}{ccccccccccccccccc} 
\tablecolumns{15} \tabletypesize{\scriptsize} \tablewidth{0pc}
\tablecaption{List of solar flares with microwave ZPs during 2000
- 2013\label{tbl-1}} \tablehead{
 \colhead{Date}& \colhead{Class}&\colhead{$t_{st}$(UT)}&\colhead{$t_{fp}$(UT)}&\colhead{$D_{ri}$(min)}&\colhead{phase}&\colhead{$N_{zp}$}&\colhead{$f_{zp}$}  & Telescope &\\}
\startdata
 2013-04-11    &  M6.5          &      06:58           &       07:16          &          18           & rising        &         4        & 2.6-3.8            &  SBRS,ORSC &\\
 2012-06-13    &  M1.1          &      12:04           &       13:15          &          71           & peak          &         3        & 1.3-1.8            &  ORSC   &\\
 2011-08-09    &  X6.9          &      08:00           &       08:04          &           4           & peak          &         1        & 2.6-3.8            &  SBRS   &\\
 2011-02-24    &  M3.5          &      07:26           &       07:35          &           9           & peak          &         2        & 2.6-3.8            &  SBRS   &\\
 2011-02-15    &  X2.2          &      01:46           &       01:56          &          10           & rising        &         1        & 5.2-7.6            &  SBRS   &\\
               &                &                      &                      &                       & decay         &         2        & 2.6-3.8            &  SBRS   &\\
 2010-08-01    &  C3.2          &      07:55           &       09:00          &          65           & rising        &         8        & 1.1-1.5            &  ORSC   &\\
 2006-12-13    &  X3.4          &      02:14           &       02:40          &          26           & rising        &         4        & 2.6-3.8            &  SBRS   &\\
               &                &                      &                      &                       & peak          &         1        & 2.6-3.8            &  SBRS   &\\
               &                &                      &                      &                       & decay         &         4        & 2.6-3.8            &  SBRS   &\\
 2006-12-05    &  X9.0          &      10:18           &       10:35          &          17           & rising        &         6        & 2.4-3.6            &  ORSC   &\\
 2005-07-11    &  C1.0          &      16:33           &       16:37          &           4           & decay         &         2        & 1.3-1.6            &  ORSC   &\\
 2005-07-09    &  M2.8          &      21:47           &       22:06          &          19           & rising        &         4        & 2.6-3.8            &  SBRS   &\\
 2005-06-01    &  C2.3          &      10:41           &       10:51          &          10           & rising        &         1        & 1.0-1.3            &  ORSC   &\\
 2005-01-16    &  X2.6          &      22:25           &       23:02          &          37           & decay         &         1        & 1.1-2.06           &  SBRS   &\\
 2005-01-15    &  X1.2          &      00:22           &       00:43          &          21           & peak          &         1        & 2.6-3.8            &  SBRS   &\\
 2005-01-15    &  M8.6          &      05:54           &       06:37          &          43           & rising        &         7        & 2.6-3.8, 1.1-1.34  &  SBRS   &\\
               &                &                      &                      &                       & peak          &         3        & 2.6-3.8, 1.1-1.34  &  SBRS   &\\
               &                &                      &                      &                       & decay         &         3        & 1.1-1.34           &  SBRS   &\\
 2004-12-02    &  M1.5          &      23:44           &       00:06          &          22           & rising        &         4        & 1.1-1.34           &  SBRS   &\\
 2004-12-01    &  M1.1          &      07:00           &       07:20          &          20           & rising        &         3        & 2.6-3.8, 1.1-1.34  &  SBRS   &\\
 2004-11-10    &  X2.5          &      01:59           &       02:13          &          14           & decay         &         1        & 2.6-3.8            &  SBRS   &\\
 2004-10-30    &  M3.7          &      09:09           &       09:28          &          19           & rising        &         2        & 2.1-3.4            &  ORSC   &\\
 2004-10-30    &  C3.7          &      12:45           &       12:51          &           6           & peak          &         1        & 2.1-2.4            &  ORSC   &\\
 2004-09-12    &  M4.8          &      00:04           &       00:56          &          52           & rising        &         9        & 2.6-3.8            &  SBRS   &\\
               &                &                      &                      &                       & peak          &         1        & 2.6-3.8            &  SBRS   &\\
 2004-05-17    &  C7.0          &      04:11           &       04:17          &           6           & peak          &         3        & 1.1-2.06           &  SBRS   &\\
 2004-04-06    &  M2.4          &      12:30           &       13:28          &          58           & peak          &         3        & 2.5-4.0            &  ORSC    &\\
 2004-01-09    &  M1.1          &      01:13           &       01:22          &           9           & peak          &         7        & 1.1-2.06           &  SBRS   &\\
 2004-01-05    &  M6.9          &      02:50           &       03:45          &          55           & rising        &         2        & 1.1-2.06           &  SBRS   &\\
 2003-11-18    &  M3.9          &      08:12           &       08:31          &          19           & rising        &         6        & 2.6-3.8            &  SBRS   &\\
 2003-10-28    &  X17           &      11:00           &       11:10          &          10           & decay         &         1        & 2.0-2.4            &  ORSC   &\\
 2003-10-26    &  X1.2          &      05:57           &       06:54          &          57           & rising        &        14        & 1.1-2.06           &  SBRS   &\\
 2003-05-29    &  M1.5          &      02:09           &       02:18          &           9           & rising        &         2        & 5.2-7.6            &  SBRS   &\\
 2003-05-27    &  M1.6          &      05:06           &       06:26          &          80           & rising        &         1        & 1.1-2.06           &  SBRS   &\\
               &                &                      &                      &                       & peak          &        12        & 1.1-2.06           &  SBRS   &\\
 2003-03-18    &  X1.5          &      11:52           &       12:08          &          16           & decay         &         1        & 2.0-2.4            &  ORSC   &\\
 2003-01-05    &  C5.8          &      05:51           &       06:17          &          26           & rising        &         2        & 5.2-7.6            &  SBRS   &\\
 2002-09-17    &  C8.8          &      09:17           &       09:21          &           4           & peak          &         5        & 2.0-4.5            &  ORSC   &\\
 2002-04-21    &  X1.5          &      00:43           &       01:10          &          27           & decay         &        12        & 2.6-3.8            &  SBRS   &\\
 2001-10-19    &  X1.6          &      00:47           &       01:05          &          18           & rising        &         5        & 2.6-3.8            &  SBRS   &\\
               &                &                      &                      &                       & decay         &         2        & 2.6-3.8            &  SBRS   &\\
 2001-09-16    &  C4.3          &      07:40           &       07:45          &           5           & rising        &         1        & 1.2-1.6            &  ORSC   &\\
 2000-11-25    &  M8.2          &      00:59           &       01:31          &          32           & rising        &        10        & 2.6-3.8            &  SBRS   &\\
 2000-11-24    &  X2.0          &      04:55           &       05:02          &           7           & rising        &         2        & 2.6-3.8            &  SBRS   &\\
               &                &                      &                      &                       & peak          &         2        & 2.6-3.8            &  SBRS   &\\
 2000-10-29    &  M4.4          &      01:28           &       01:57          &          29           & decay         &        14        & 2.6-3.8            &  SBRS   &\\
 2000-06-06    &  X2.3          &      15:00           &       15:26          &          26           & decay         &        14        & 2.0-3.5            &  ORSC   &\\
 2000-04-09    &  M3.1          &      23:26           &       23:42          &          16           & rising        &         1        & 2.6-3.8            &  SBRS   &\\\hline
 Sum           &  40 flares     &                      &                      &                       &               &       202        &                    &         &\\
\enddata

\tablecomments{$Class$: the GOES soft X-ray class, $t_{st}$: flare
start time, $t_{fp}$: flare peak time, $D_{ri}$: flare rising
time, $t_{zp}$: the central time of ZP; $f_{zp}$: frequency range
of ZP (GHz); $N_{zp}$: number of ZP events in the flare. SBRS
indicates the Chinese Solar Broadband Radio Spectrometers at
Huairou, and ORSC indicates the Ond\'rejov radiospectrograph in
Czech Republic.}
\end{deluxetable}

In order to investigate the relationship between ZP and the
flaring processes, we define a phase time $P_{ph}$ to describe the
relative time of ZP structure occurrence with respect to the
maximum of solar flare:

\begin{equation}
P_{ph}=(t_{zp}-t_{fp})/(t_{fp}-t_{st}).
\end{equation}

$t_{zp}$ is the central time of ZP structure. $t_{fp}$ and
$t_{st}$ are the peak and start time of the GOES soft X-ray (SXR)
flare, respectively. $t_{st}$ is defined when four consecutive 1
minutes SXR values have met all three of the following conditions:
(1) all four values are above the background threshold, (2) all
four values are strictly monotonically increasing, and (3) the
last value is 1.4 times greater than the value occurred 3 minutes
earlier. $D_{ri}=t_{fp}-t_{st}$ is the flare rising time.
$P_{ph}=-1.00$ indicates the time at flare start, $P_{ph}=0$
indicates the flare maximum (peak time), and $P_{ph}>0$ indicates
the time after the flare maximum. According to $P_{ph}$, the
flaring process can be partitioned into three phases:

(1) Rising phase: $P_{ph}\leq -0.25$, SXR intensity is increasing
rapidly. During this phase, the flaring region may undergo
continuously magnetic flux emergence, reconnections, energy
releasing, and plasma heating.

(2) Peak phase: $-0.25<P_{ph}\leq 0.25$, SXR intensity is
relatively stable, and has only a slightly variation. During this
phase, the magnetic flux emergence and the magnetic energy
releasing in the flaring region may reach to a steady state.

(3) Decay phase: $P_{ph}> 0.25$, SXR intensity is decreasing
slowly and continuously. During this phase, besides some local
small scale magnetic reconnections, the main energy releasing is
ended, and the flaring region may undergo a process of thermal
dissipations and cooling.

In left panel of Figure 1, the dotted curve is an example profile
of GOES SXR emission in a typical flare, the vertical dashed lines
partition flaring process into rising, peak, and decay phases.

\begin{deluxetable}{ccccccccccccccccc} 
\tablecolumns{10} \tabletypesize{\scriptsize} \tablewidth{0pc}
\tablecaption{The parameters of microwave ZPs in a M8.6 flare on
2005 January 15\label{tbl-1}} \tablehead{
 \colhead{Phase}& \colhead{$t_{zp}$}&\colhead{$P_{ph}$} &\colhead{$N_{str}$}&\colhead{$f_{zp}$(MHZ)} &\colhead{$\Delta f$}(MHz)&\colhead{$\Delta f/f_{zp}$(\%)}&\colhead{$r(\%)$} & \colhead{$D_{zp}$}(s)&\\}
\startdata
  rising        &  06:13:10         & -0.55             &   5               &       1204             &            28           &              2.33             &      100         &          0.5          \\
                &  06:13:27         & -0.54             &   5               &       1296             &            36           &              2.78             &      100         &          4.0          \\
                &  06:14:58         & -0.51             &   2               &       2680             &           100           &              3.73             &        0         &          5.0          \\
                &  06:15:20         & -0.50             &   5               &       1192             &            26           &              2.18             &      100         &          2.0          \\
                &  06:16:32         & -0.48             &   2               &       2730             &            90           &              3.30             &        0         &          3.0          \\
                &  06:23:02         & -0.32             &   5               &       1196             &            18           &              1.51             &      100         &          0.8          \\
                &  06:24:34         & -0.29             &   4               &       1196             &            20           &              1.67             &      100         &          3.5          \\\hline
  peak          &  06:27:39         & -0.22             &   3               &       2750             &           105           &              3.82             &        0         &          4.0          \\
                &  06:28:36         & -0.22             &   3               &       2750             &           120           &              4.36             &        0         &          4.5          \\
                &  06:31:44         & -0.12             &   6               &       1240             &            52           &              6.77             &        0         &         30.0          \\\hline
  decay         &  06:49:10         &  0.28             &   5               &       1224             &            48           &              3.92             &       20         &         45.0          \\
                &  06:51:07         &  0.33             &   9               &       1180             &            18           &              1.53             &       85         &          0.8          \\
                &  07:17:32         &  1.18             &   8               &       1150             &            14           &              1.22             &      100         &          3.3          \\
\enddata
\tablecomments{$t_{zp}$: central time (UT); $P_{ph}$: phase time,
$f_{zp}$: central frequency; $\Delta f$: frequency separation of
zebra stripes, $\Delta f/f_{zp}$: relative frequency separation,
$N_{str}$: zebra stripe number, $D_{zp}$: ZP duration, $r$:
averaged polarization degree.}
\end{deluxetable}

Besides the phase time $P_{ph}$, we also collect the ZP central
frequency ($f_{zp}$), stripe number ($N_{zp}$), frequency
separation between adjacent zebra stripes ($\Delta f$), relative
frequency separation ($\Delta f/f_{zp}$), averaged degree of
polarization ($r$), and the ZP duration ($D_{zp}$). Table 2 lists
the parameters of ZPs occurring in a long-duration flare. In this
work, we measured all above parameters of each ZP event in the
sample, which will be analyzed in the following sections.

\section{Statistical Results and Analysis}

Based on the above sample of 202 microwave ZP events, we present
comprehensively statistical investigations in this section.

\subsection{The ZP dependence with flares}

At first, we hope to know which kind of flares and what phase of
the flare may be preferential to produce ZP phenomena. Table 1
indicates that among the 40 flares accompanying with ZPs, there
are 23 flares having ZPs at rising phase, 14 flares having ZPs at
peak phase, and 12 flares having ZPs at decay phase. There are
only 2 flares having ZPs at all three phases (an M8.6 flare on
2005 January 15, and an X3.4 flare on 2006 December 13).

Table 1 also shows that there are 14 X-class flares with averaged
rising time $D_{ri}$ 25.5 min, 18 M-class flares with averaged
rising time $D_{ri}$ 31.7 min, and 8 C-class flares averaged
rising time $D_{ri}$ 15.8 min accompanying with microwave ZPs.
During the same observing period, the whole numbers of X, M, and
C-class flares observed by the two instruments are 75, 805, and
1330, respectively, and their averaged rising times are 16.5 min,
14.8 min, and 11.5 min, respectively. Additionally, as for the
flares accompanying with microwave ZPs, the correlation
coefficient between the flare's rising time and the ZP number is
0.56, which implies that they have significant correlation. The
comparison of these numerical values imply that flares with longer
rising time are preferential to produce microwave ZPs. At the same
time, the powerful flares are more preferential to produce
microwave ZPs than the relatively weak flares.

The left panel of Figure 1 presents the ZP distribution with
respect to the phase time in flares, which shows that the
microwave ZP can occur in all rising, peak, and decay phases of
flares. Among the 202 microwave ZPs, there are 96 (47.5\%)
occurred in the flare rising phase, 50 (24.7\%) occurred in the
flare peak phase, and 56 (27.8\%) occurred in the flare decay
phase. The contour profile of the ZP distribution in Figure 1
implies that microwave ZPs are more preferential to produce before
the flare maximum (64.9\%).

\subsection{The parameter properties of ZPs}

(1) ZP Frequency distribution

Among the 202 ZPs, there are 72 ZPs with central frequency in the
range of 1.00 - 2.00 GHz, 87 ZPs with central frequency in the
range of 2.00 - 3.00 GHz, 37 with central frequency in the range
of 3.00 - 4.00 GHz, and only 6 ZPs with central frequency above
4.00 GHz. However, as there is a difference of the observation
time between different frequency range, we make a generalization
on the above statistical values by the time lengths of the
instrument observations at each frequency domain. Additionally, as
the cadence of ORSC during 2000 - 2005 is 100 ms, which is much
longer than after 2006 and SBRS/huairou, we multiply a weight
factor 0.5 to its time length during the corresponding period,
empirically. The right panel of Figure 1 is the ZP distribution
with respect to the emission frequency, which shows that almost
90\% ZPs are occurring below 4.00 GHz, and 2.00 - 3.00 GHz is the
most preferential frequency domain to produce ZP structures.

\begin{figure*}[ht] 
\begin{center}
   \includegraphics[width=8 cm]{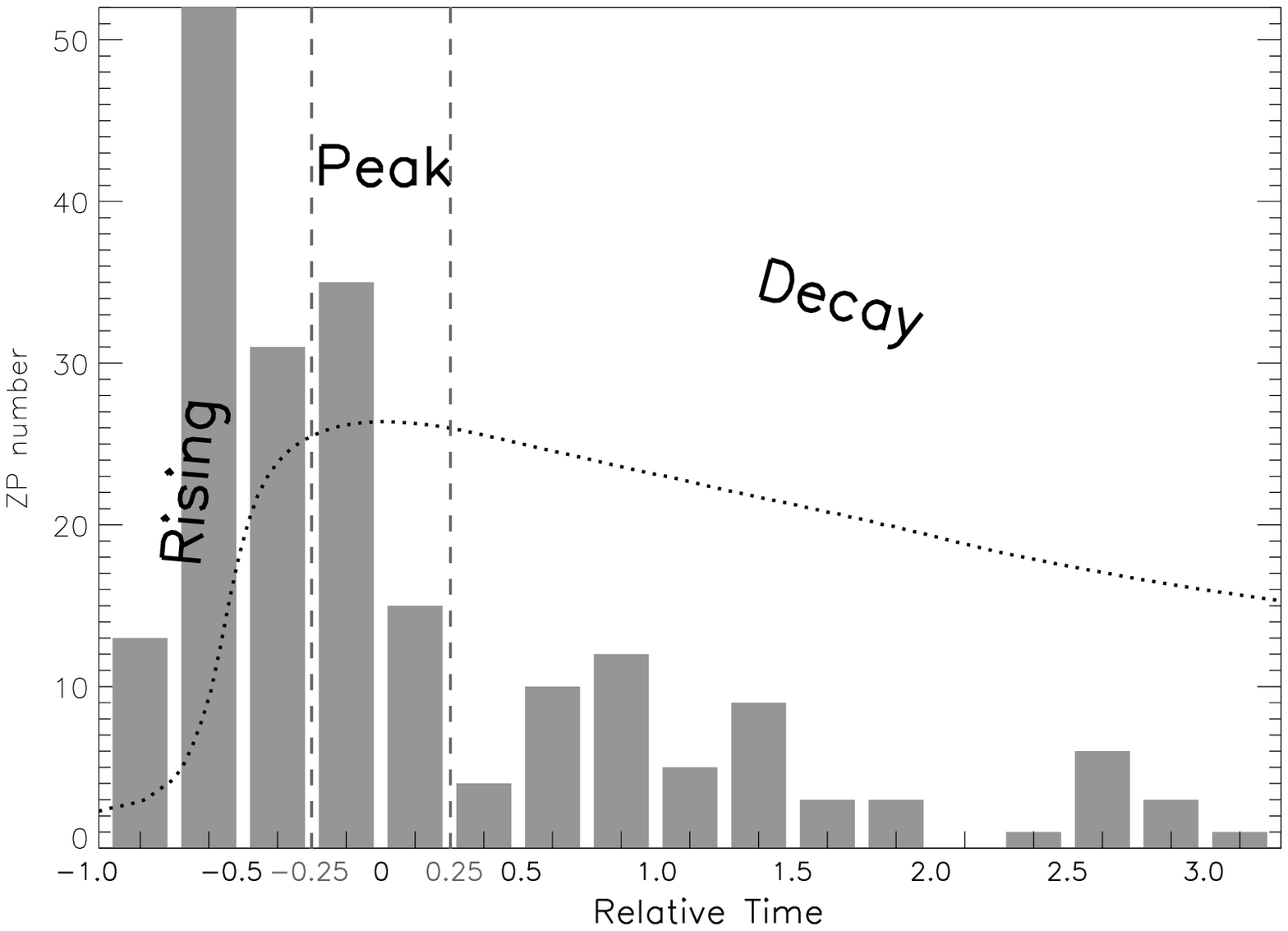}
   \includegraphics[width=8 cm]{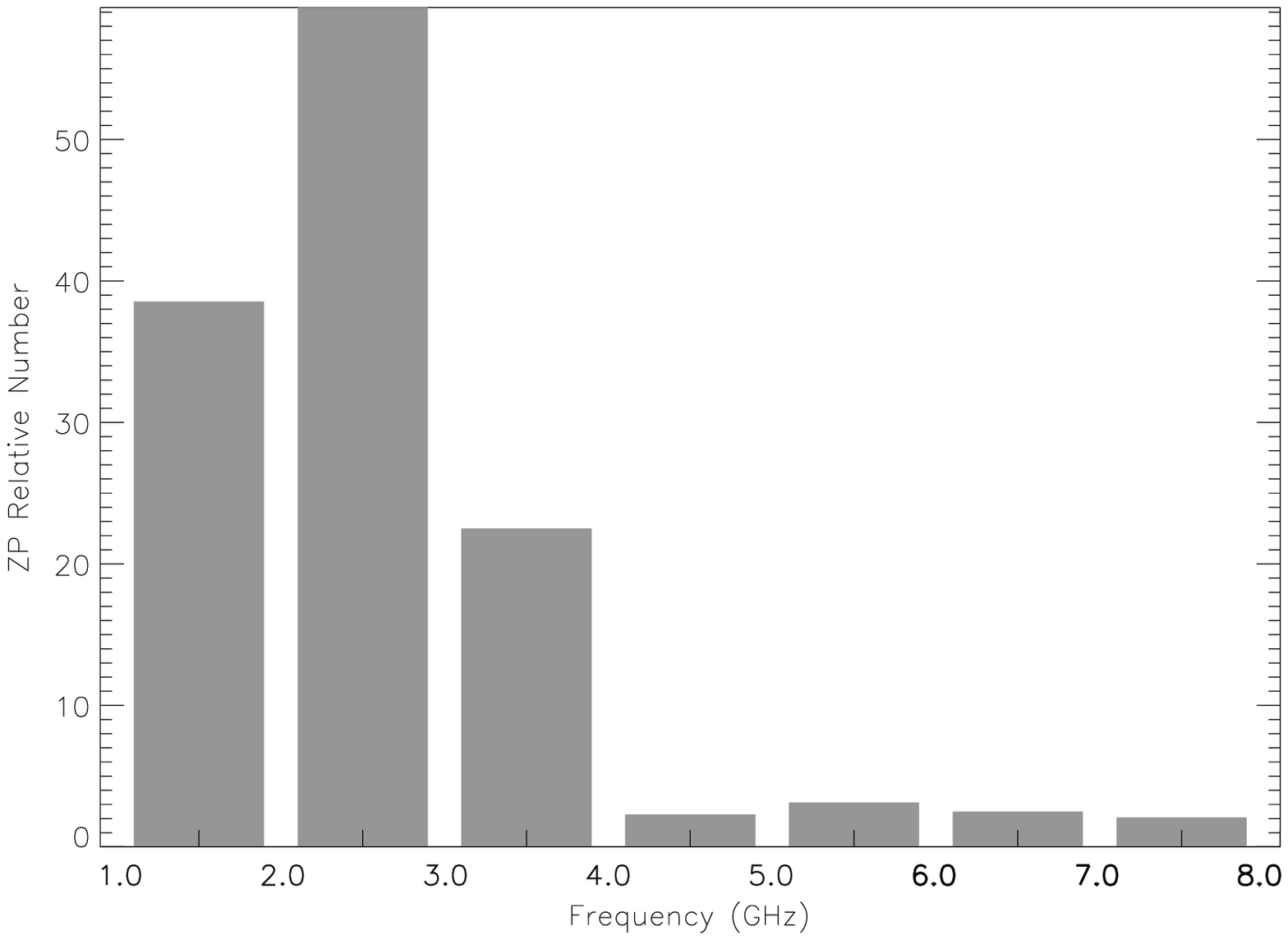}
\caption{Left is the ZP distribution with respect to the phase
time in flares. The dotted curve is an example profile of GOES
soft X-ray emission in a typical flare. The two vertical dashed
lines partition flaring process into rising, peak, and decay
phases. Right is the ZP distribution with respect to the emission
frequency.}
\end{center}
\end{figure*}

(2) Polarization

Generally, the polarization sense of spectral fine structures is
an important parameter. From Table 2, we may find that the
microwave ZPs in a same flare may have almost all polarization
modes from very weak (near 0) to very strong (near 100\% at LCP or
RCP). As there is no polarization observation from ORSC, here, we
only analyze the polarization properties of the 154 ZPs obtained
by SBRS/Huairou. Among these 154 ZP events, there are 68 (44.2\%)
ZPs have strong polarization degree ($r\geq 80\%$), 38 (24.7\%)
ZPs have moderate polarization degree ($20\%<r<80\%$), and 48
(31.1\%) ZPs have no obviously polarizations ($r<20\%$). This fact
indicates that there is no dominated polarization modes in
microwave ZPs.

(3) Duration

Table 2 indicates that the duration of ZPs even in same flare are
also distributed in a wide range from sub-second to several
decades seconds. The left-lower panel of Figure 2 presents the
distribution of ZP durations with respect to the relative time of
ZP occurred in the flares, which shows that ZP duration has a
moderate linear increase from the flare early rising phase to its
late decay phase. The dash-dotted line is a result of chi-square
goodness linear fitting. Although the distribution is very
disperse in the flare rising and peak phases. Maybe the
statistical averaged values of the ZP duration can give some
supplementary inspirations. In the flare rising phase, the
averaged ZP duration is 5.41 s with variance 8.85 s, and the
relative variation is 1.64. Around the flare peak phase, the
averaged ZP duration is 6.95 s with variance 11.03 s, and the
relative variation is 1.59. While in the flare decay phase the
averaged ZP duration is 17.55 s with variance 21.44 s, and the
relative variation is 1.22, which is much smaller than that in the
other two phases and in the whole sample.

\begin{figure*}[ht] 
\begin{center}
   \includegraphics[width=8 cm]{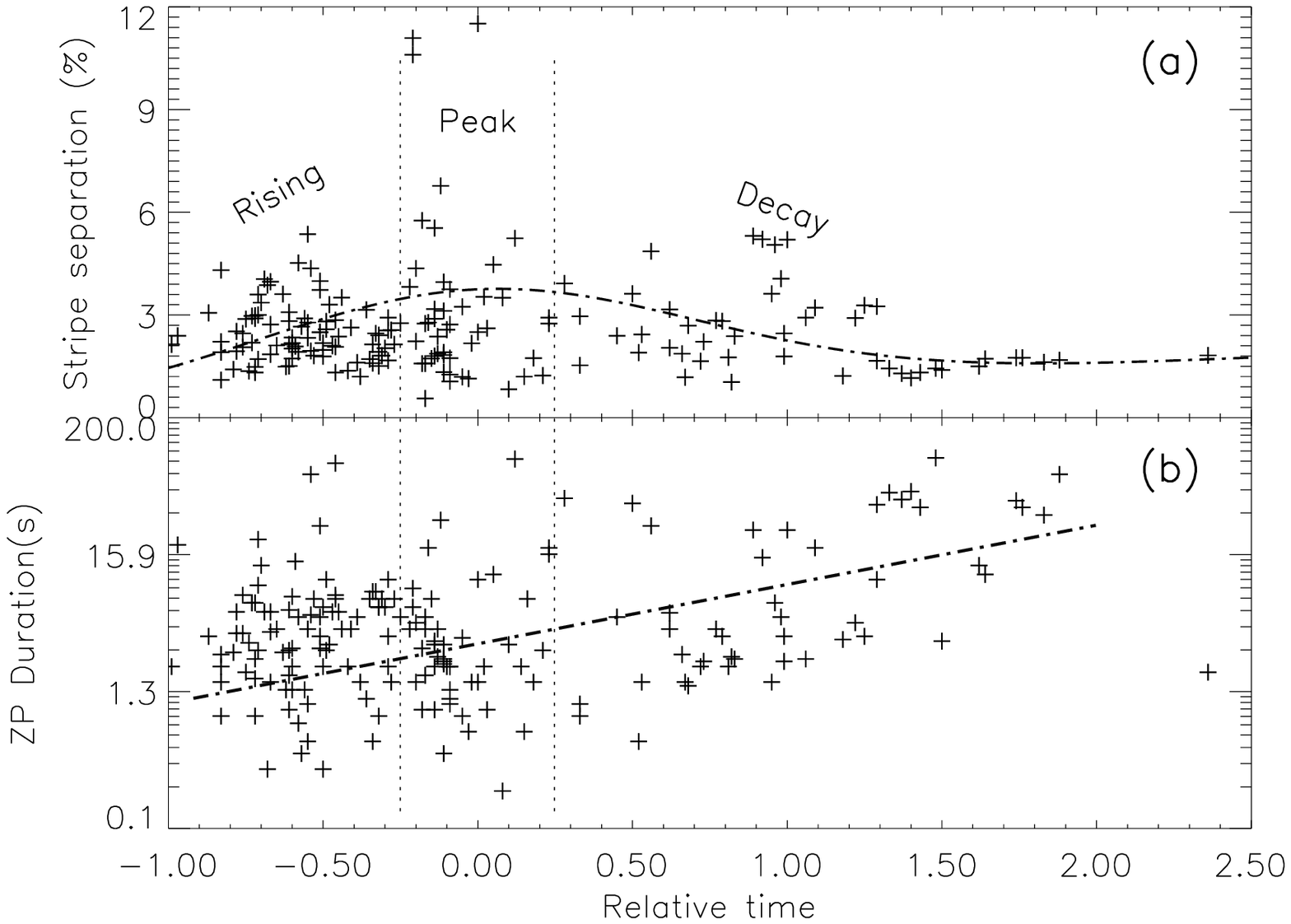}
   \includegraphics[width=8 cm]{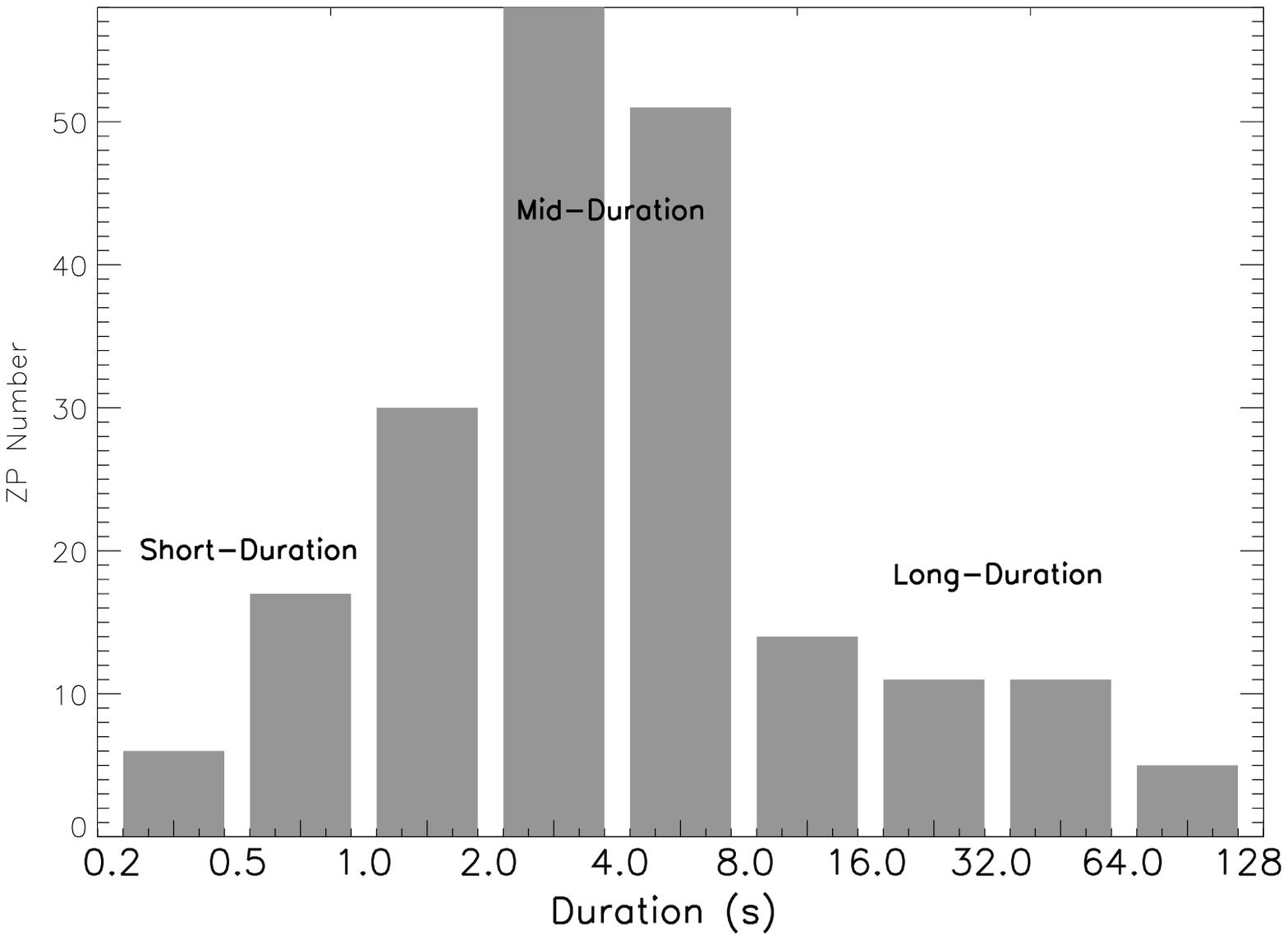}
\caption{Left is the distributions of zebra stripe frequency
separation (a) and durations (b) with respect to the phase time of
ZP occurring. Right is the generalized ZP distribution with
respect to the durations.}
\end{center}
\end{figure*}

Actually, the minimum duration in the whole sample (with 202 ZP
events) is only 0.2 s, while the longest duration is 95 s, which
covers about 3 orders of magnitude. The right panel of Figure 2
presents a histogram of the distribution of ZPs numbers with
respect to ZP durations. Here, we set the interval scale of the ZP
duration in power exponent of 2. It is showed that there are 153
ZPs (75.7\%) having the duration in range of 1 - 16 s, while only
23 ZPs (11.4\%) have durations $<1.0$ s and 26 (12.9\%) ZPs have
durations $>16.0$ s. This fact implies that the very short and
very long duration ZPs are very rare.

(4) Frequency separation of zebra stripes

The frequency separations between the adjacent zebra stripes are
in the range of 14 - 340 MHz. It depends on the ZP central
frequency. Generally, the higher ZP central frequency, the wider
of the frequency separations among zebra stripes. We define a
relative separation: $\Delta f/f_{zp}$, here, $\Delta f$ and
$f_{zp}$ are the zebra stripe frequency separation and the ZP
central frequency, respectively. The sixth and seventh column of
Table 2 listed the $\Delta f$ and $\Delta f/f_{zp}$ in a typical
flare, which indicates that $\Delta f/f_{zp}$ is in the range of
several percents. In fact, among the whole sample, the maximum of
$\Delta f/f_{zp}$ can be high up to 10\%, while the minimum can be
low down to below 1\%. Statistical calculation indicates that
averaged $\Delta f/f_{zp}$ is 2.42\% in the flare rising phase,
3.23\% in the flare peak phase, and 2.49\% in the flare decay
phase. This variation can be fitted by an exponential curve,
showed in the left-bottom panel of Figure 2 (the dash-dotted
curve), although it is very scattered in a broad range around the
flare peak.

It is very interesting to investigate the variations of the
frequency separation with respect to its frequency in each ZP
event. In order to make such investigation, we just analyze the ZP
events with at least 4 zebra stripes (there are at least 3 values
of the frequency separation), and totally there are 151 ZP events
occurring in 33 flares in our sample. Here, we find that there
exist three kinds of obviously different variations of the
frequency separation:

(1) Constant separation, the amplitude of $\triangle f$ variation
does not exceed the frequency resolution of the spectrometer,
which can be approximately regarded as a constant. A1 panel of
Figure 3 is an example of this kind, here, the $\triangle f$
variation is 4 MHz, which is just the frequency resolution of the
telescope.

(2) Varying separation, the amplitude of $\triangle f$ variation
exceeds the frequency resolution of the spectrometer, and is
scattered in a wide range. A2 panel of Figure 3 is an example of
this kind, here, the total variation of $\triangle f$ is 44 MHz
which is 11 times of the frequency resolution (4 Mhz) and has
changes in addition-and-deletion.

(3) Rising separation, the amplitude of $\triangle f$ variation
also exceeds the frequency resolution of the spectrometer and
increases continuously with respect to its frequency. A3 panel of
Figure 3 is an example of such kind.

\section{Classification of ZPs}

It is meaningful to make a physical classification of the
microwave ZPs, which may help us to understand the basic
properties of their origin and applications to solar eruptions.
However, so far, there is no such work in the existing
publications. The main reason is that it is very difficult to
collect enough microwave ZP events to form a considerable big
sample for a reasonable physical classification. The statistical
analysis in above sections indicates that parameters of emission
frequency, stripe frequency separation, polarization degree,
stripe number, and the phase time in the associated flare are
always distributed dispersively in wide ranges. For example, the
ZPs occurred in flare rising phase may have all polarization
degrees from very weak ($r\rightarrow0$) to very strong
($r\rightarrow100\%$) circular polarized modes as well as in flare
peak phase or decay phase. The most of the other parameters also
have the similar dispersive characteristics.

\begin{figure*}[ht] 
  \includegraphics[width=5.4 cm]{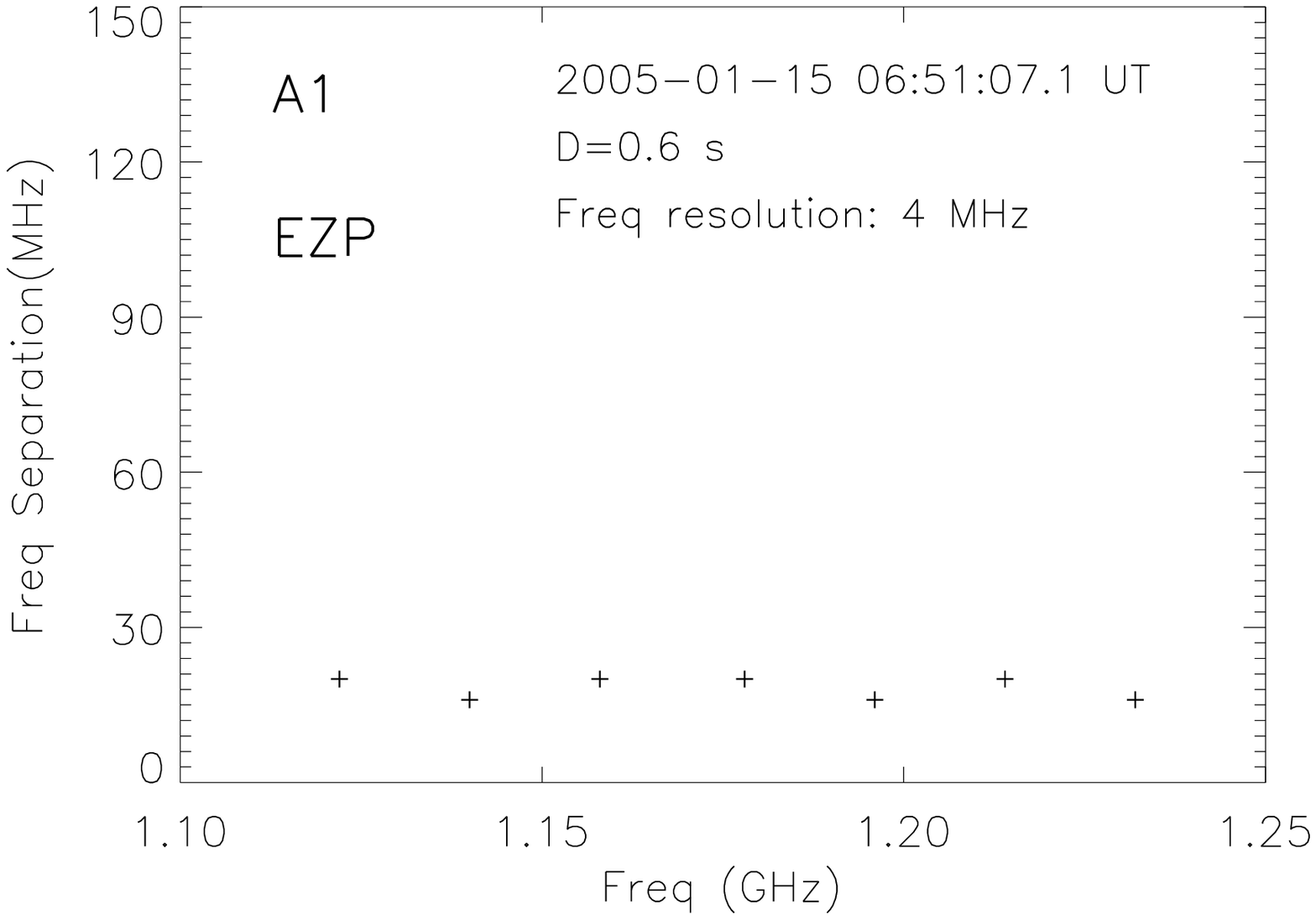}
  \includegraphics[width=5.4 cm]{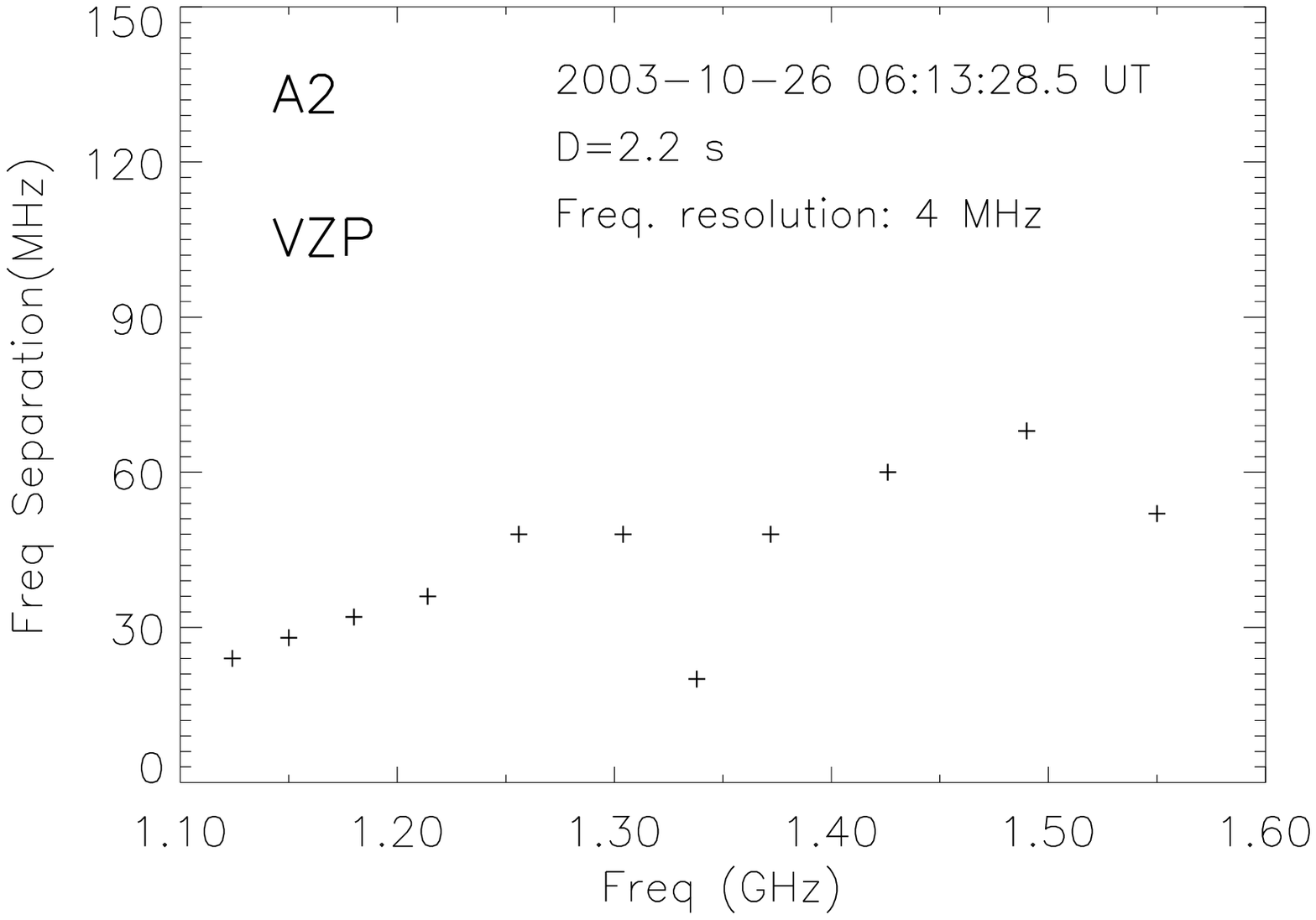}
  \includegraphics[width=5.4 cm]{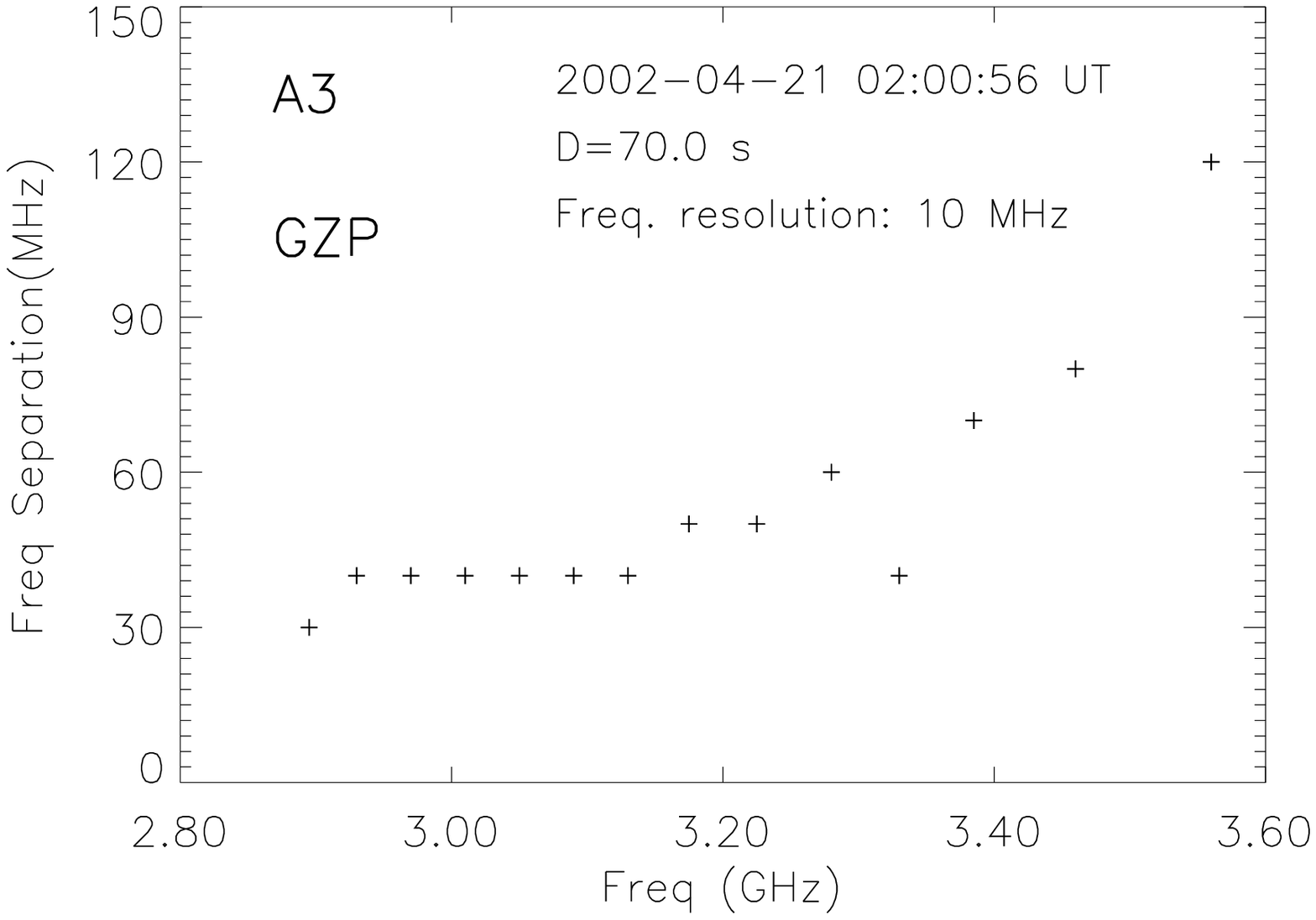}
  \includegraphics[width=5.4 cm]{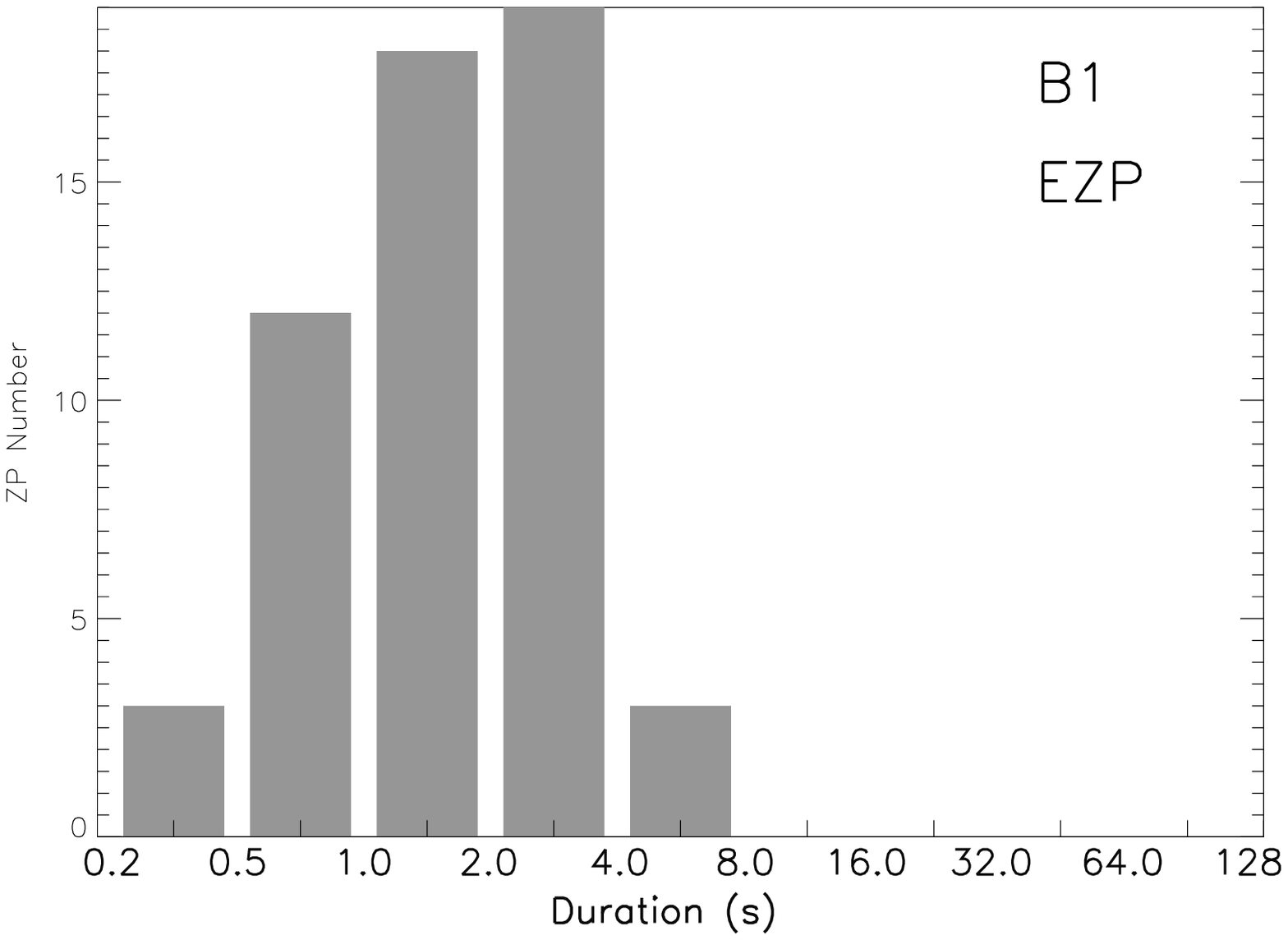}
  \includegraphics[width=5.4 cm]{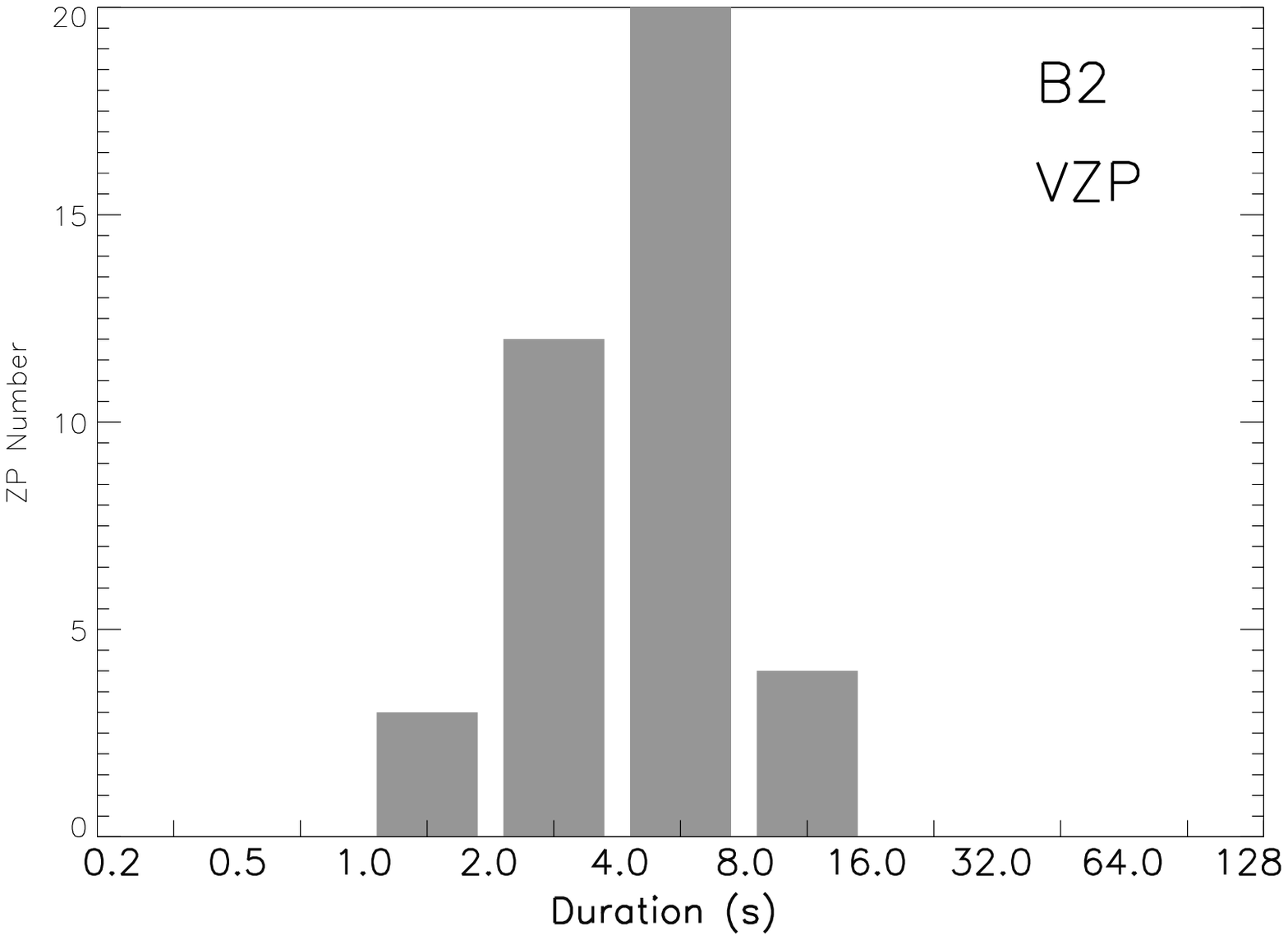}
  \includegraphics[width=5.4 cm]{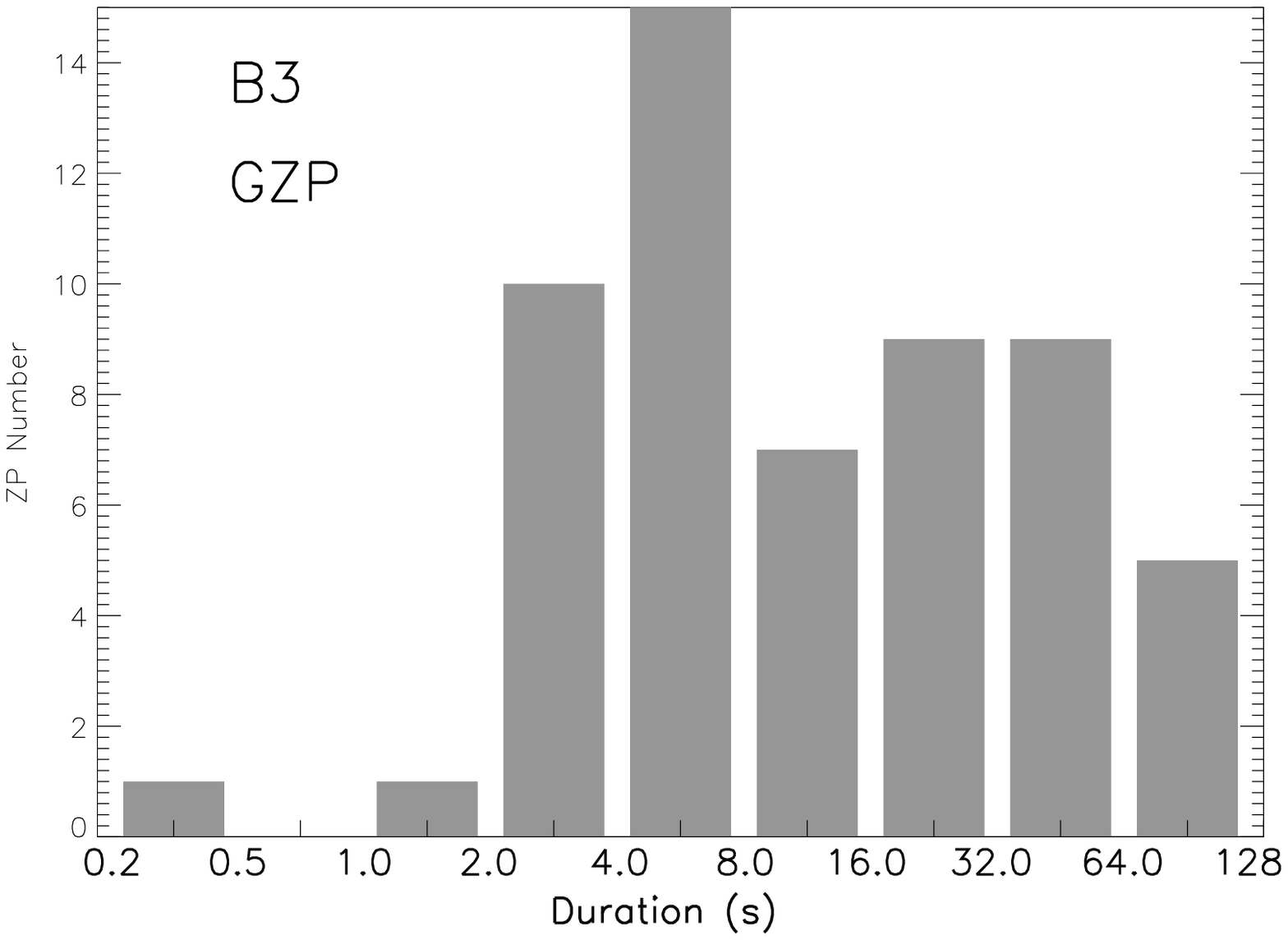}
  \includegraphics[width=5.4 cm]{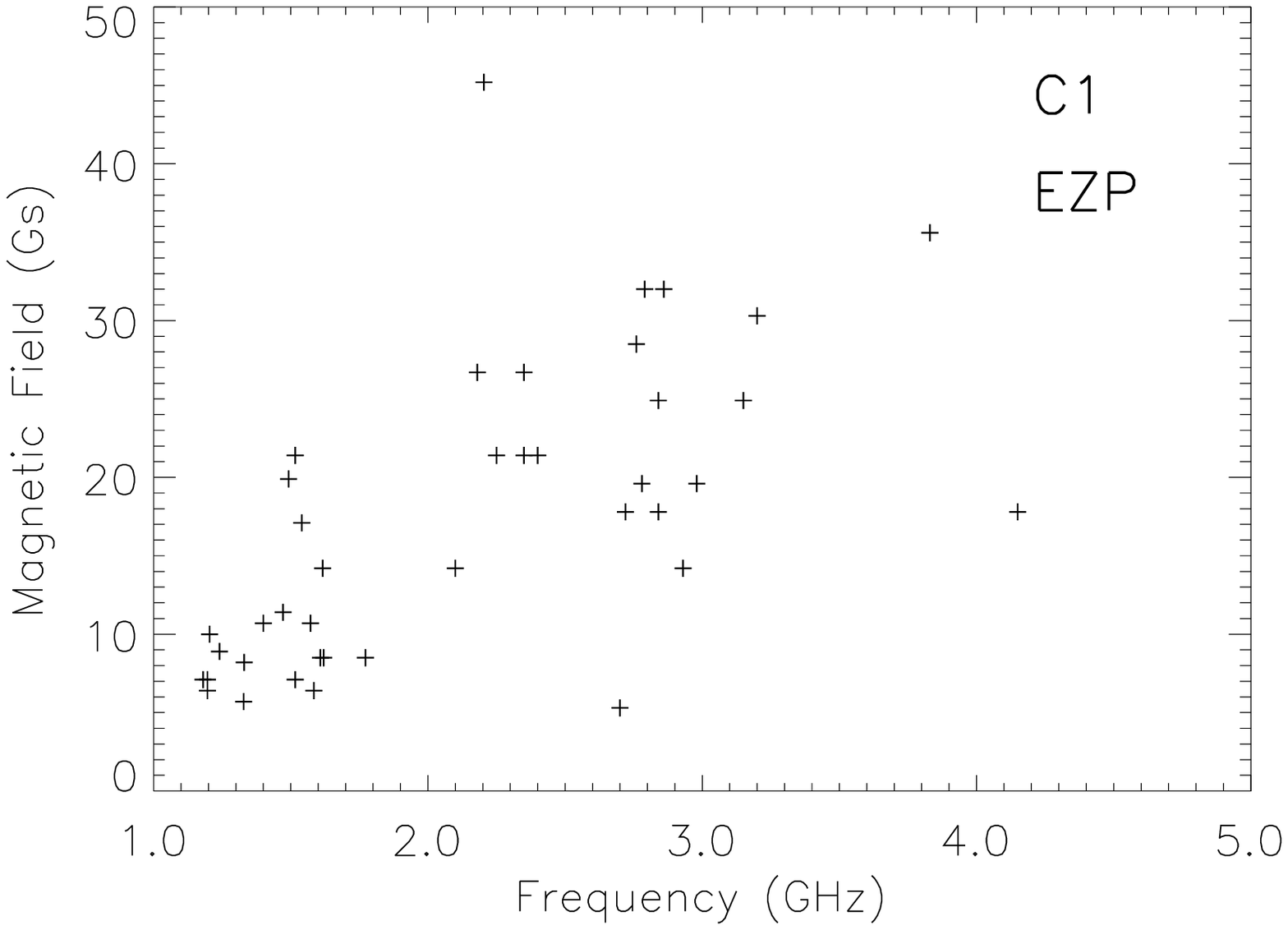}
  \includegraphics[width=5.4 cm]{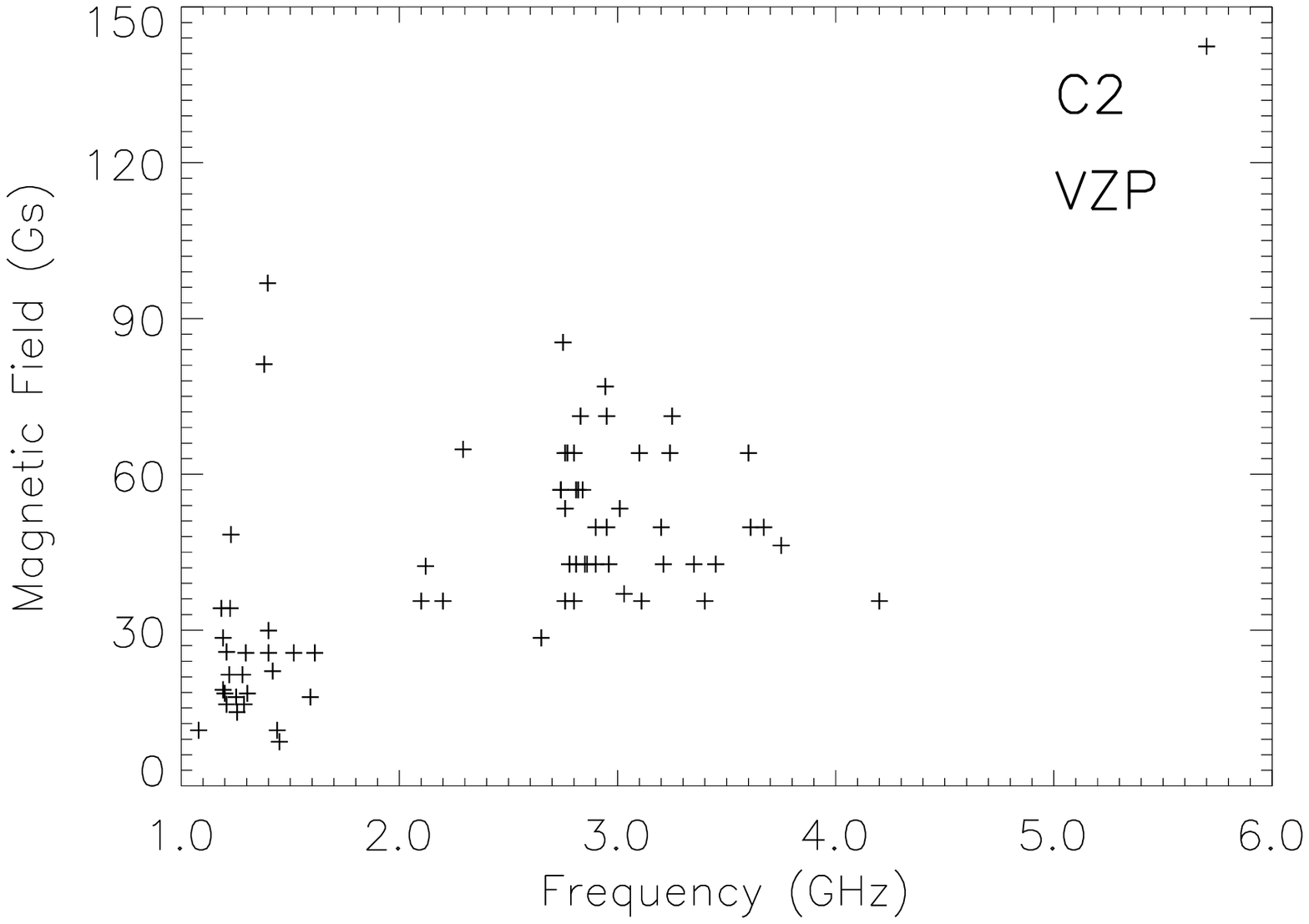}
  \includegraphics[width=5.4 cm]{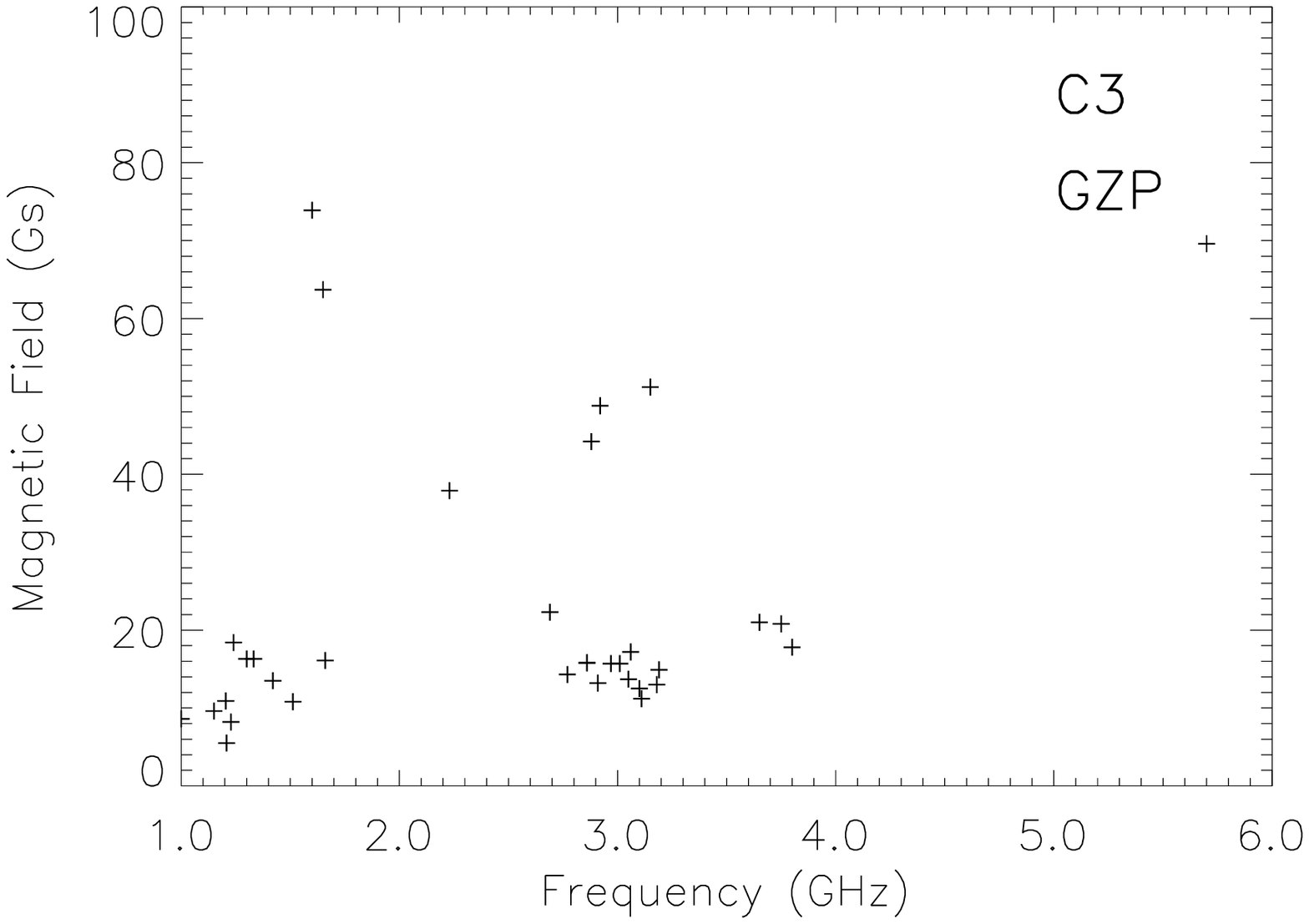}
  \includegraphics[width=5.4 cm, height=6.5 cm]{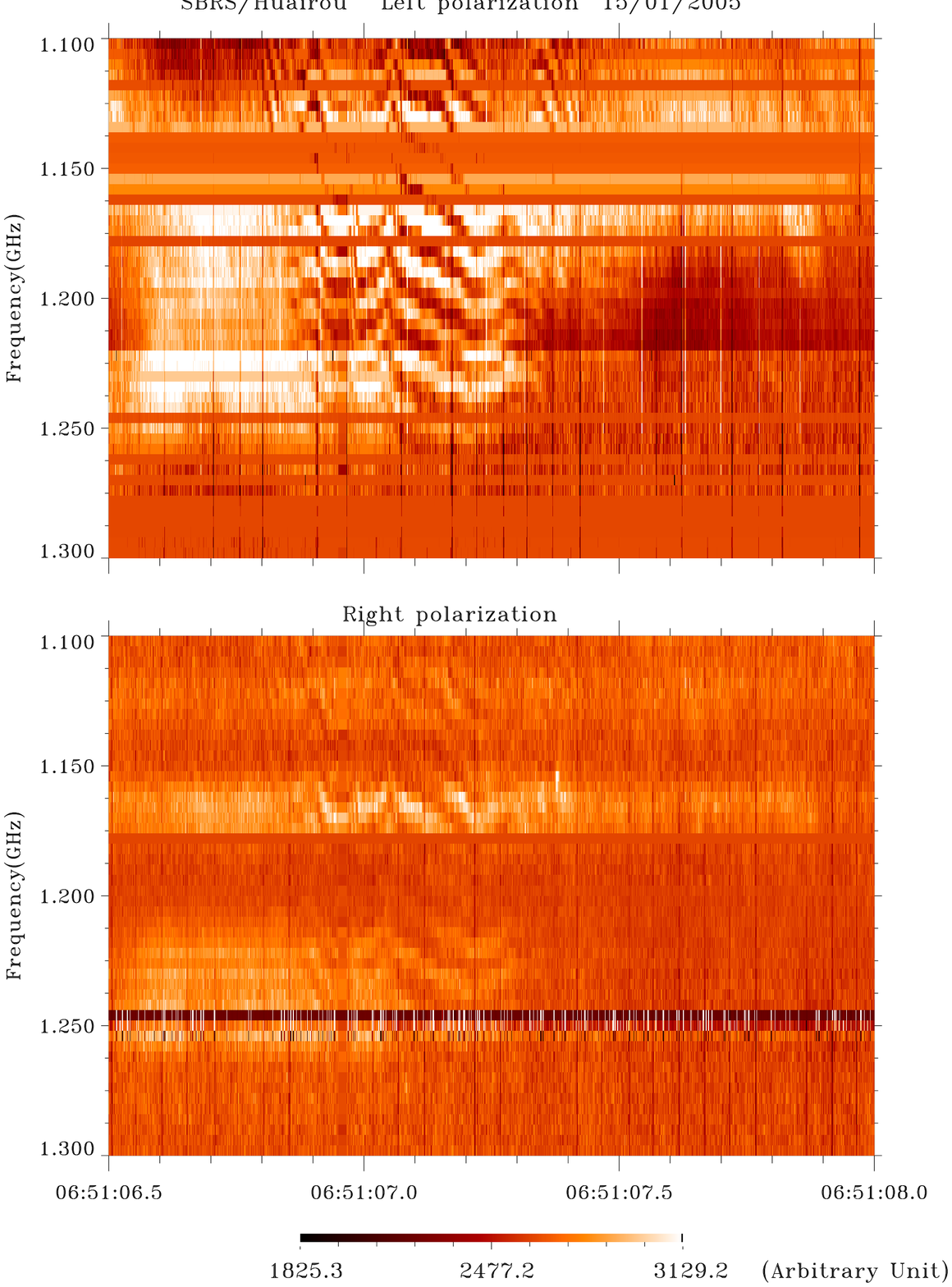}
  \includegraphics[width=5.4 cm, height=6.5 cm]{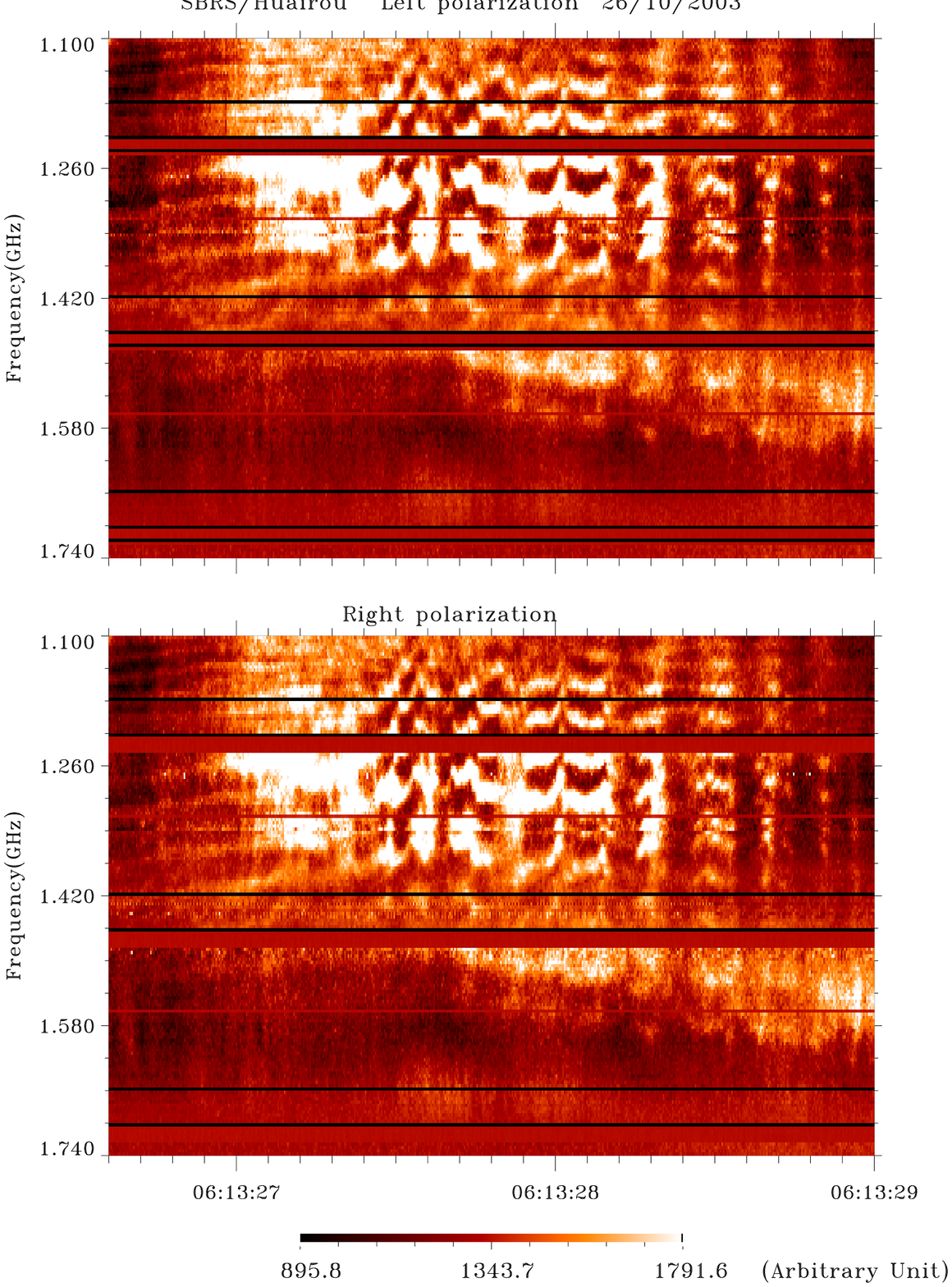}
  \includegraphics[width=5.4 cm, height=6.5 cm]{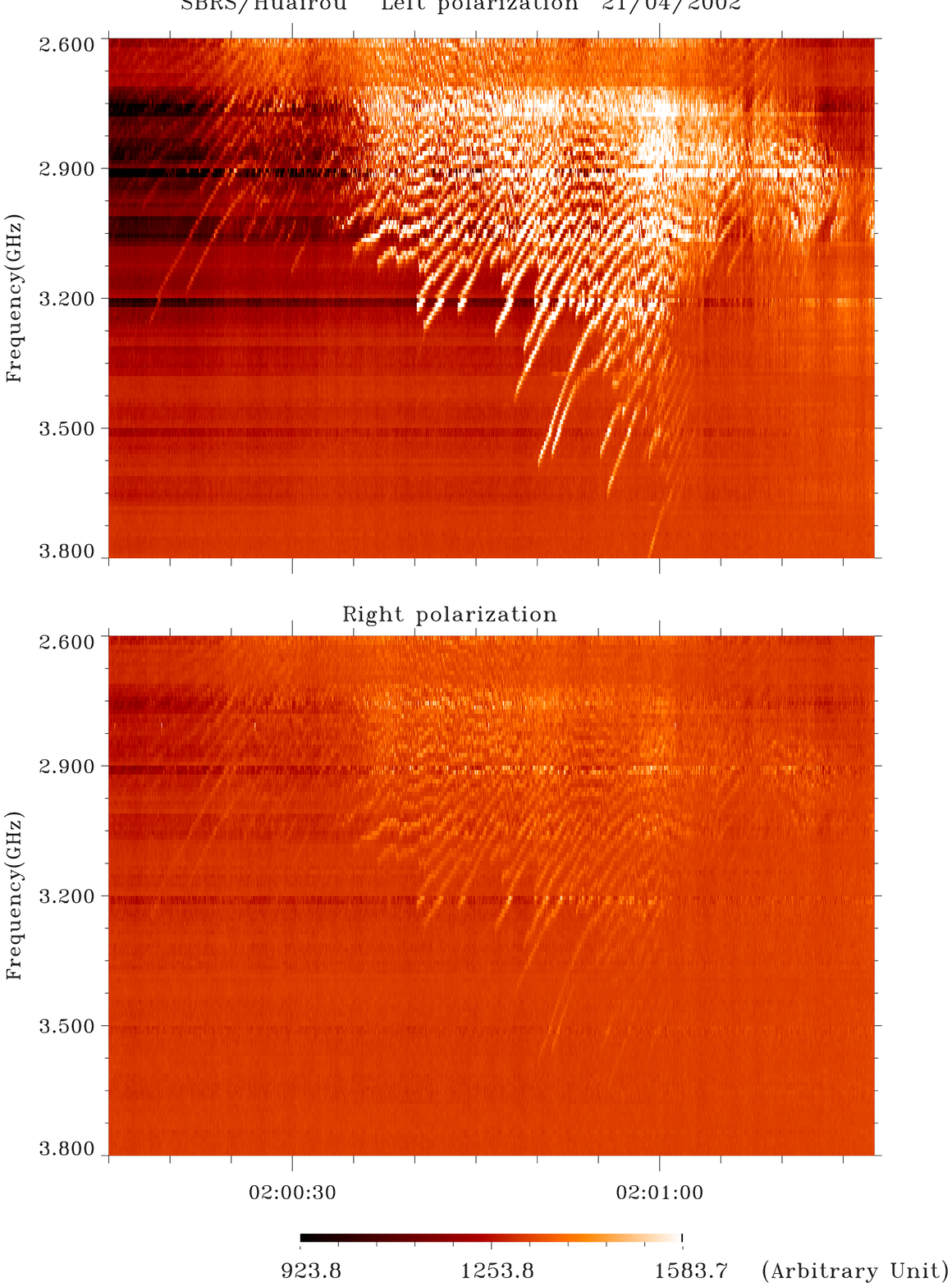}
\caption{A1, A2, and A3 show the variations of the frequency
separation between adjacent zebra stripes versus the stripe number
from low to high frequency in typical equidistant ZP,
variable-distant ZP, and growing-distant ZP; B1, B2, and B3 show
the ZP distributions versus durations in these three ZP types; C1,
C2, and C3 are the magnetic fields in the source regions deduced
from each type of ZPs, respectively. The bottom panels are the
spectrograms of typical equidistant ZP (left) with constant
separation, variable-distant ZP (middle) with varying separation
and growing-distant ZP (right) with rising separation,
respectively.} \label{fig:source}
\end{figure*}

However, the above analysis implies that it is very valuable when
we combine duration and the variation of the frequency separation
with respect to its frequency in each ZP event. According to these
two factors, we may classify ZPs into three types with other
parameters having relatively narrow range. The detailed properties
can be presented as following:

Equidistant ZP (EZP), which stripe frequency separation
approximates to a constant. Among the 151 ZPs with more than 4
stripes, there are 55 ZPs belonging to EZP, and their distribution
to the duration is showed in B1 of Figure 3. Additionally, we find
that (1) the durations of EZP is ranging from 0.3 s to 5.0 s, and
most of them have durations around 1 - 2 s, the average duration
is 1.6 s. (2) most EZPs are strong circular polarization, there
are 33 short-duration ZPs observed by SBRS/Huairou which show that
there are 20 events with strong circular polarizations, the
averaged polarization degree $r\sim 70\%$. (3) most EZPs occurred
in the flare rising and peak phases, only 7 events occurred at the
flare decay phase, and the averaged $P_{ph}=-0.31$. (4) The zebra
stripe number is less than 10 with average value of about 6.5. (5)
Most EZPs are simple and isolated far away from the other spectral
fine structures. The bottom left panel of Figure 3 is an example
of EZP, which is very simple and isolated ZP without any other
fine spectral structures before and after it. The duration is only
0.6 s, with strong left-handed circular polarization.

Variable-distant ZP (VZP), which stripe frequency separation
varies in a wide range and with irregularly increasing or
decreasing. Among the 151 ZPs, there are 39 ZPs belonging to VZP,
and their distribution to the duration is showed in B2 of Figure
3. Additionally, we find that (1) the duration of VZP is in the
range of 1.5 - 14 s, and most of them are around 2 - 10 s, the
averaged value is 6.4 s. (2) The polarization degree has a very
wide range from 0 to 100\%. (3) VZPs can appear in flare rising,
peak phases as well as in flare decay phase. (4) The zebra stripe
number of VZP is also in a wide range of 2 - 28, and most of them
are less than 10 stripes. (5) most VZPs are relatively complex,
and always accompanying with other fine structures. The bottom
middle panel of Figure 3 is a typical example of VZP, which is
occurred in the rising phase of an X1.2 flare on 2003 October 26.
The duration is 2.2 s with weakly polarization. It is a complex ZP
modulated by a quasi-periodic pulsation with period of about 100
ms, similar to the quasi-periodic wiggles modulated by some MHD
oscillations (Yu et al. 2013).

Growing-distant ZP (GZP), which stripe frequency separation has a
big variation and increases continuously with respect to its
frequency. Among the 151 ZPs, there are 57 ZPs belonging to GZPs,
and their distribution to duration is showed in B3 of Figure 3.
Additionally, we find that (1) their duration ranges from 0.2 s to
95 s, most of them are longer than 10 s, the average duration is
27.6 s. (2) most GZPs are moderate or very weak circular
polarizations. Among these 57 GZPs, there are 35 events having
polarization degree $r<50\%$. (3) most GZPs occurred in the flare
decay phase. Totally, there are 41 ZPs occurring in flare decay
phase, therein 25 ZPs belong to GZP. (4) most GZPs have more than
15 zebra stripes, the average stripe number in each ZP is 14.9,
which is much more than that occurring in EZPs and VZPs. (5) most
GZPs are very complex and accompanied or superimposed with many
other spectral fine structures, such as fibers, spikes, narrow
band type III bursts, fast quasi-periodic pulsations, etc. The
bottom right panel of Figure 3 is a typical example of GZP, which
is occurred in the very deep decay phase of a long-duration X1.5
flare on 2002 April 21. The ZP duration is about 70 s, besides its
left-handed circular polarization can be distinguished bright
zebra stripes, the right-handed circular polarization can be also
identified zebra stripes clearly. It is a very complex ZP
accompanying with many other spectral fine structures. On the
whole, it is a combination of ZP structure with quasi-periodic
wiggles at the relatively low frequency side and a fiber structure
at the relatively high frequency side. On the details, each zebra
stripe are consisting of superfine millisecond spikes (Chernov et
al. 2005, Chen \& Yan 2007).

Figure 3 can present a clearly comparison among the three kinds of
ZP types. Actually, the comparison indicates that the above
classification of microwave ZPs can present the different physical
processes with each other to some extent, although there are a bit
of overlapping between EZP and VZP, or between VZP and GZP at
durations. Sometimes, there may be a transition from EZP to VZP,
or from VZP to GZP in a same ZP structure.

As we know that BW model induced that the frequency separation
between the adjacent zebra stripes is a constant and approximates
the electron gyro-frequency $f_{ce}$, and all stripes may produce
from a small compact source region. The corresponding duration
will be very short, and the resulting spectrum has only a few
harmonics (less than 10). These items conform with the basic
characteristics of the EZPs. Therefore, BW model possibly reveals
the basic mechanism of EZPs, and the stripe frequency separation
can directly measure the magnetic field in the source region:
$B\approx 3.56\times 10^{-7}\Delta f$. Here, the unit of $B$ is
Gs, and $\Delta f$ in Hz. C1 panel of Figure 3 presents the
magnetic fields in the source regions deduced from EZPs. Here, we
find that the magnetic field strength ranges mainly from 10 Gs to
45 Gs, and has an increasing trend with emission frequencies.

DPR model proposes that zebra stripes is produced from some
resonance levels where the upper hybrid frequency coincides with
the harmonics of electron gyro-frequency in the inhomogeneous flux
tube, and the frequency separation is dominated not only by the
electron gyro-frequency, but also by the gradient of plasma
density. Since the DPR levels present in the non-uniform trap and
the kinetic instability can be excited by a small quantity of
trapped non-equilibrium electrons, the DPR mechanism can provide a
fairly large number of stripes (e.g., more than 20 stripes) in the
ZP spectra with comparatively long durations. Based on the common
value of magnetic field and plasma density around the flaring
source region, we know that the frequency separation will have a
slowly increase with respect to the frequency. These facts conform
with the main characteristics of GZPs, which shows that DPR model
may explain their formation. By using DPR model, we can also
deduce the magnetic fields in the source region: $B\approx
3.56\times 10^{-7}Q\cdot\Delta f$,
$Q=\frac{1}{n}|\frac{2H_{p}}{H_{b}}-1|$, when polarization is
strong, $n=1$, and when polarization is weak, $n=2$. C3 panel of
Figure 3 is the distribution of magnetic field strength with
respect to emission frequency deduced from GZPs, the magnetic
field strength ranges from 10 Gs to 75 Gs, which is more dispersed
than that of EZPs and VZPs.

It seems very difficult to make a reasonable explanation to the
formation of VZPs for their irregular variation of frequency
separations. WW model induced that the zebra stripes frequency
separation is about: $\Delta f=2f_{w}$, and the whistler wave
frequency: $f_{w}\approx 0.1 - 0.5 f_{ce}$, here, $f_{ce}\ll
f_{pe}$. Therefore, $\Delta f$ will have variations in a
relatively narrow band. These properties seem to indicate that the
WW model should be the possible mechanism to explain the formation
of VZPs. Applying WW model, an approximated estimation of the
magnetic field in the source region can be obtained. As we know
that the whistler wave group velocity peaks at frequency
$\frac{1}{4}f_{ce}$, which indicates that the frequency separation
$\Delta f$ may vary around $\frac{1}{2}f_{ce}$. Then we may
estimate the magnetic field strength just by the frequency at
whistler peak group velocity: $B\approx 7.12\times 10^{-7}\Delta
f$. C2 panel of Figure 3 is the distribution of magnetic field
strength with respect to emission frequency deduced from VZPs by
above method. Here, we find that the magnetic field strength
ranges from 10 Gs to 145 Gs, which is stronger and much wider
distribution than that in the source of EZPs. However, as we know
that the WW model has many problems which have not been proved so
far. There are many work need to study comprehensively. The
diversity of polarization sense of VZPs imply that it is also
possible that the VZP may be a blended spectral structure produced
from some combined mechanisms (for example, the propagating model,
or the combination of DPR and WW models, etc.).

Because the BW mechanism requires a larger number of
non-equilibrium electrons than that of DPR mechanism, and there
are more non-equilibrium electrons in the flare rising phase for
the continuously magnetic reconnection than that in the flare
decay phase, BW mechanism may be more preferential to work in the
flare early phase and small source region to produce EZPs. The DPR
mechanism may be more preferential to work in the flare decay
phase and produce GZPs from different resonance levels in
relatively stable flaring loop. In the flare decay phase, there
are many small scale magnetic reconnections and energy releases in
the hot magnetized plasma loops, many small scale microwave bursts
(Tan 2013) may take place accompanying with microwave ZPs, such as
microwave spikes, narrow band type III bursts, and fast
quasi-periodic pulsations or wiggles (Tan et al. 2007, Yu et al.
2013). In WW mechanism, the low-frequency whistler waves are
excited by non-equilibrium electrons with loss-cone distributions
in coronal traps with intermittent non-uniform layers (Chernov
2006), such conditions may appear in all phases of solar flares,
and therefore VZPs may take place in all flaring phases.

\section{Conclusions and Discussions}

There are many parameters which apply to describe the
characteristics of microwave ZPs associated with solar flares,
such as the central frequency ($f_{zp}$), phase time ($P_{ph}$),
polarization degree ($r$), zebra stripe number ($N_{str}$),
duration ($D_{zp}$), frequency separation between adjacent zebra
stripes ($\Delta f$) and the relative value ($\Delta f/f$).
However, from the statistical investigation, we find that most
parameters can not act as the classifying indicator of microwave
ZPs, while the combination of duration and the variation of the
frequency separation with respect to its frequency in each ZP
event may provide a physical classification. With such
combination, we may classify the microwave ZPs into three types:

(1) EZP, simple and isolated with constant frequency separation of
the adjacent zebra stripes, very short duration ($1.0 - 2.0 s$),
relatively strong polarization, less than 10 stripes, and mainly
occurred in the flare rising phase.

(2) VZP, relatively complex with irregular varying frequency
separation of the adjacent zebra stripes and mid-term durations
($2.0 - 10.0 s$), diverse polarization modes, and always
overlapped by some other structures, such as quasi-periodic
pulsations or wiggles, etc.

(3) GZP, very complex with increasing frequency separation of the
adjacent zebra stripes and long duration ($>10 s$), relatively
weak polarization, and mainly occurred in the flare decay phase
with more than 10 stripes and accompanying with many spectral fine
structures, such as spikes, fibers, and quasi-periodic pulsations.

Different types of microwave ZPs may have different formation
mechanisms, and therefore may reveal different physical processes
in the source regions. The main properties of EZP indicate that
the BW model should be the best mechanism to explain its
formation. VZPs may produce from WW wave mechanism or from some
complex multi-mechanisms. And the DPR model may reveal the
physical processes of GZPs. The estimation of the magnetic field
strengths deduced from the above models and ZP structures shows
that the magnetic field in the microwave ZP source regions ranges
from 10 Gs to 145 Gs, which is in the acceptable domain of the
magnetic field in coronal flaring source regions.

The above classification may help us to clarify the controversies
among the existing various ZP models. Of course, since the
theories of BW model, WW model, and DPR model are far from perfect
(Zlotnik 2009). Many physical details are still not clear, we do
not know exactly which model is the best one to explain the
formation of a given ZP event. For example, it is difficult to
distinguish whether a ZP with only three or less zebra stripes
belongs to EZP, or VZP, even or GZP. We have to look for other
properties of ZPs and further theoretical and observational
investigations. The another problem is the formation of VZPs for
their diversity and irregularity, it is possible that it is formed
from some complex mechanism. Additionally, it is still a big
problem why some zebra stripes are composed of many millisecond
spikes with super-high brightness temperature? And what is the
physical relationship between ZP structure and its inner
millisecond spikes? These questions need us to study more
comprehensively, especially the observational information with
high spatial resolutions at the corresponding frequencies.

From the statistical analysis, it is found that microwave ZPs can
occur in the flare rising and peak phases as well as in the flare
decay phase, especially more preferential to produce in
long-duration powerful flares around frequency of 3.00 GHz. Such
fact implies that there are some common characters attached in
microwave ZPs. As we know the microwave emission source region
around 3.00 GHz is possibly very closed to the core region of
solar flaring and energy releasing, microwave ZPs may reveal some
fundamental nature of solar eruptive processes.

\acknowledgments The authors would like to thank the referee for
the helpful and valuable comments on this paper. We would also
thank the the GOES, NoRP, ORSC/Ond\'rejov, and SBRS/Huairou teams
for providing observation data. This work is supported by NSFC
Grant 11273030, 11221063, 11373039, 11103044, MOST Grant
2011CB811401, the National Major Scientific Equipment R\&D Project
ZDYZ2009-3, and the Grant P209/12/00103 (GA CR). This work was
also supported by the Marie Curie PIRSES-GA-295272-RADIOSUN
project.

\end{document}